\documentclass[manuscript,authorversion,nonacm]{acmart}
\newcommand{\quotes}[1]{``#1''} 
\usepackage{paralist}
\usepackage{sistyle} 
\usepackage{booktabs} 
\newcommand{\tabitem}{~~\llap{\textbullet}~~} 
\usepackage{cleveref}
\usepackage{tabularx} 
\newcolumntype{L}{>{\raggedright\arraybackslash}X} 
\usepackage{multicol} 
\usepackage{multirow} 
\usepackage{graphicx} 
\usepackage[export]{adjustbox} 
\usepackage{makecell} 
\usepackage{pifont} 
\newcommand{\cmark}{\ding{51}}
\usepackage{cleveref} 
\usepackage{adjustbox} 
\newcommand{\autorefappendix}[1]{\hyperref[#1]{Appendix~\ref*{#1}}}  
\usepackage{footnote}
\usepackage[shortlabels]{enumitem} 
\usepackage[font=small,labelfont=bf]{caption} 
\usepackage{dblfloatfix} 
\usepackage{wrapfig}
\usepackage{sistyle} 
\SIthousandsep{,}
\usepackage[edges]{forest}
\usetikzlibrary{arrows.meta,shadows.blur}
\usepackage{pdflscape}
\usepackage{tcolorbox}
\def\fillandplacepagenumber{%
 \par\pagestyle{empty}%
 \vbox to 0pt{\vss}\vfill
 \vbox to 0pt{\baselineskip0pt
   \hbox to\linewidth{\hss}%
   \baselineskip\footskip
   \hbox to\linewidth{%
     \hfil\thepage\hfil}\vss}}
\makeatletter
\def\@copyrightspace{\relax}
\makeatother

\begin{document}



\title{Privacy Perceptions and Behaviors of Google Personal Account Holders in Saudi Arabia}

\author{Eman Alashwali}
\authornote{Eman Alashwali was a Collaborating Visitor at CMU while working on this paper.}
\affiliation{%
  \institution{King Abdulaziz University (KAU) and King Abdullah University of Science and Technology (KAUST)}
  \country{Saudi Arabia}
  }
\email{ealashwali@kau.edu.sa}

 \author{Lorrie Cranor}
\affiliation{%
  \institution{Carnegie Mellon University (CMU)}
  \country{United States}
  }
\email{lorrie@cmu.edu}
\begin{tcolorbox}
This is the author's manuscript for an extended version of a paper appeared in Proc. of the 26th International Conference on Human-Computer Interaction (HCII), 2024 available at: \url{https://link.springer.com/chapter/10.1007/978-3-031-61379-1_1} 
\end{tcolorbox}
\begin{abstract}
While privacy perceptions and behaviors have been investigated in Western societies, little is known about these issues in non-Western societies. To bridge this gap, we interviewed 30 Google personal account holders in Saudi Arabia about their privacy perceptions and behaviors regarding the activity data that Google saves about them. Our study focuses on Google's Activity Controls, which enable users to control whether, and how, Google saves their Web \& App Activity, Location History, and YouTube History. Our results show that although most participants have some level of awareness about Google's data practices and the Activity Controls, many have only vague awareness, and the majority have not used the available controls. When participants viewed their saved activity data, many were surprised by what had been saved. While many participants find Google's use of their data to improve the services provided to them acceptable, the majority find the use of their data for ad purposes unacceptable. We observe that our Saudi participants exhibit similar trends and patterns in privacy awareness, attitudes, preferences, concerns, and behaviors to what has been found in studies in the US. Our results emphasize the need for: \begin{inparaenum}[1)] \item improved techniques to inform users about privacy settings during account sign-up, to remind users about their settings, and to raise awareness about privacy settings; \item improved privacy setting interfaces to reduce the costs that deter many users from changing the settings; and \item further research to explore privacy concerns in non-Western cultures\end{inparaenum}.
\end{abstract}

\maketitle

\thispagestyle{plain}
\pagestyle{plain}

\section{Introduction} \label{sec:intro}
Web and mobile application users share a vast amount of their data with multiple parties, including the service provider, other users, and third parties. Although many applications include privacy settings, these settings often have fairly permissive defaults (i.e., sharing users' data by default)~\cite{chen19,haselton17}, with controls that are confusing, even for security and privacy experts~\cite{germain22,green18}. Users who are unaware of default settings and how to change them may share data accidentally~\cite{liu11,madejski12,shih15,green18},  sometimes with consequences such as embarrassment, bullying, identity theft, stalking, and fraud~\cite{gross05}. In some cases, regulators have accused service providers of misleading users with privacy setting interfaces. For example, in 2022 Google agreed to a \$392 million settlement in the US for misleading consumers with privacy setting interfaces that failed to clearly inform users about how to turn off location tracking~\cite{kang22}. Previous work has investigated users' perceptions and behaviors with respect to privacy settings in various contexts, such as social media~\cite{gross05,ellison07,liu11,madejski12,habib22}, smart home devices~\cite{malkin19,zheng18,lau18}, and mobile apps~\cite{ramokapane19,felt12}. However, little published research has investigated privacy settings of web services' accounts such as Google, or privacy settings in non-Western societies. 

Our study aims to address both of these gaps by exploring Saudi users' privacy perceptions (awareness, attitudes, preferences, and concerns) and behaviors, regarding the data Google saves about them. We focus on Google's Activity Controls, a section in Google accounts that enables users to control whether, and how, Google saves their Web \& App Activity, Location History, and YouTube History. To this end, we interviewed 30 Google personal account holders in Saudi Arabia. Our systematic qualitative analysis allowed us to identify fine-grained themes. Our key findings can be summarized as follows:
\begin{itemize} 
 \item Most participants have some level of awareness about Google's data practices and the Activity Controls, but many have only vague awareness. Only a few participants reported they became aware of Google's Activity Controls when signing up for their Google accounts. 
 
 \item Most participants have not actually used Google's Activity Controls. The cost of knowledge, attention, time, memory, and the fear of messing things up deterred many users from using them. 
 
\item Many participants expressed negative sentiments after they viewed their saved activity data in their Google accounts because they felt watched and were surprised by the extent of the data that had been saved. 

\item Most participants are opposed to the use of their activity data for showing ads on Google's websites and apps or on Google's partners' websites and apps. On the other hand, most of them find it acceptable for Google to use their data for improving its services.


\item Most of the participants who plan to take future steps to protect their privacy said they are going to adjust their Google Activity Controls to more restrictive settings, such as turning on the Auto-delete and reducing the data retention periods. 
 
\item We observe that our Saudi participants exhibit similar trends and patterns in privacy awareness, attitudes, preferences, concerns, and behaviors to what has been found in studies in the US. However, our study is not a replication of any of the US studies, and further research is needed to directly compare US and Saudi participants.
\end{itemize}    

Based on our results, we emphasize the need for: \begin{inparaenum} \item improved techniques to inform users about privacy settings during  account sign-up, to remind users about their settings, and to raise awareness about privacy settings; \item improved privacy setting interfaces to reduce the costs that deter many users from changing the settings; and \item further research to explore privacy concerns in non-Western cultures. \end{inparaenum}

\section{Background and Related Work} \label{sec:related}
In~\autoref{sec:account}, we introduce the Google account sign-up process, the Activity Controls, and related work on Google's Activity Controls. In~\autoref{sec:priv_settings_issues}, we summarize related research conducted in Western societies (in the US, unless stated otherwise) on users' privacy perceptions and behaviors towards privacy settings and data practices. In~\autoref{sec:saudi}, we provide a brief background on Saudi Arabia, and related work on privacy issues for Saudi users. 
 \subsection{Google Account Sign-Up and the Activity Controls} \label{sec:account}
\begin{figure}
	\begin{center}
	\includegraphics[width=0.50\textwidth]{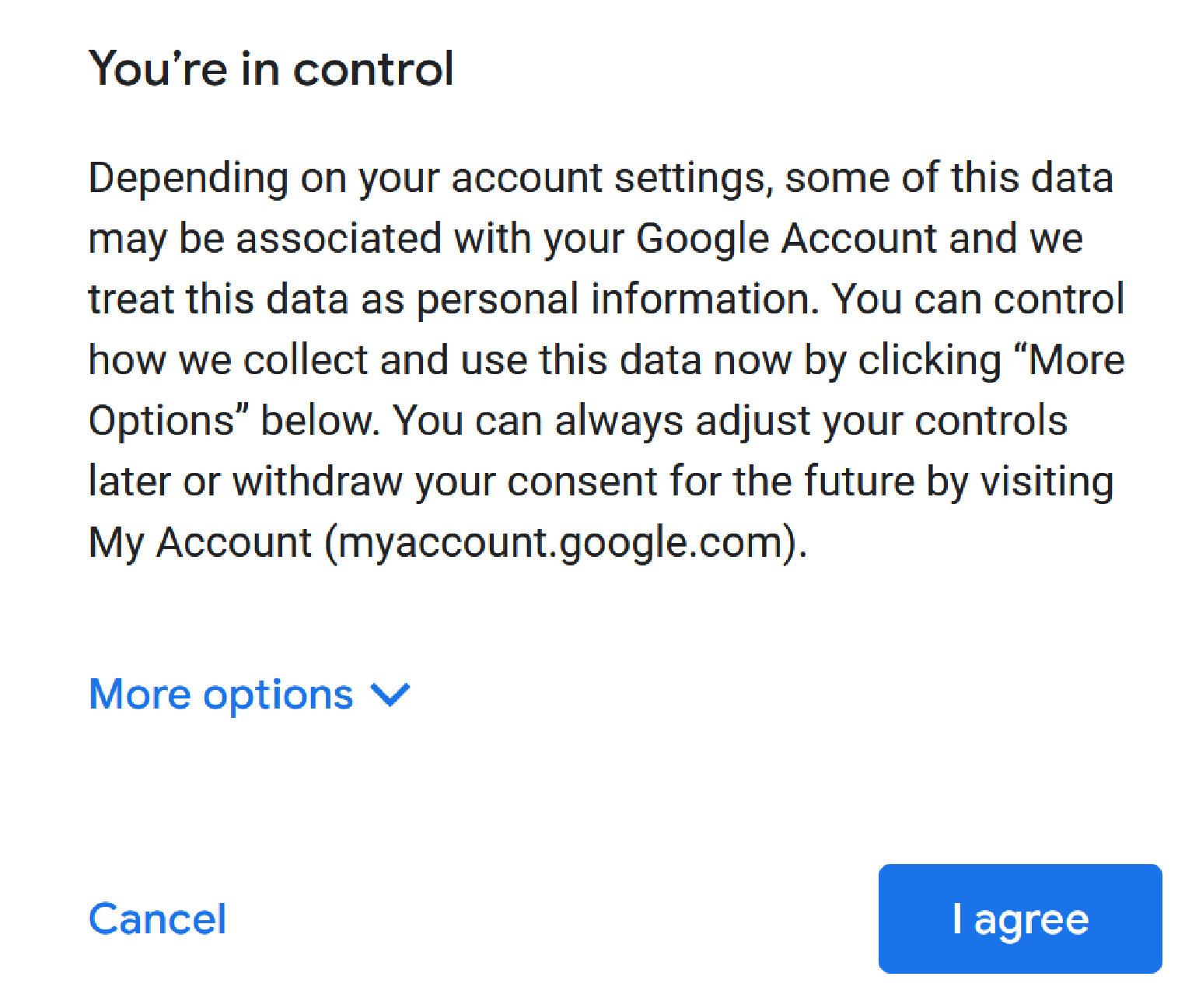}
    \end{center}
	\caption{Google's \quotes{Privacy and Terms} page which appears during the sign-up process. Only if the \quotes{More options} link is clicked, users can change the hidden permissive default privacy settings.}
	\label{fig:options}
 \end{figure}

Google is one of the largest companies in the world. As of 2018, Google reported 1.5 billion active Gmail users~\cite{kerns18}. When a user creates a free Google account (Gmail), Google presents the user with it's Privacy and Terms page. The last section entitled \quotes{You're in control} informs the user that \quotes{You can control how we collect and use this data by clicking `More options' below.}  The page presents a blue \quotes{I agree} button and links to \quotes{Cancel} and \quotes{More Options}  (see~\autoref{fig:options} for an illustration). Interfaces with this design are often described as \quotes{dark patterns,} or \quotes{deceptive patterns,} as they manipulate users to choose settings that they may not choose otherwise~\cite{brignull}. In Google's case, only if the user clicks on the \quotes{More options} link,
the page expands and displays privacy settings and the following explanation: \quotes{Customize your activity on Google sites and apps, including searches and associated info like location ....} The page shows the settings for the following: Web \& App Activity, Ads Personalization, and YouTube History. The settings are binary, and enabled by default (permissive). For example, the Web \& App Activity options are: \quotes{Save my Web \& app activity in my Google account} and \quotes{Don't save my Web \& app activity in my Google account,} with the first checked by default. At the end of the page, there is a check box for: \quotes{Send me occasional reminders about these settings,} which is unchecked by default. 

We are aware of two studies that examined Google's Activity Controls with US participants, investigating different research questions than ours. Farke et al. conducted a pre-post-study, with some similar aspects to our study, to examine the effect of exposing users to the then-called Google's My Activity on their privacy perceptions and behaviors~\cite{farke21}. In a study very different from ours, Ul Haque et al. investigated how trust toward service providers and usage of Google's Activity Controls vary based on users' technical literacy~\cite{haque23}. 

\subsection{Privacy Perceptions and Behaviors Towards Privacy Settings and Data Practices}
\label{sec:priv_settings_issues}
Heyman et al. defined two privacy concepts: \quotes{privacy as subject,} which refers to privacy between users, and \quotes{privacy as object,} which refers to privacy between a user and a business~\cite{heyman14}. For simplicity, we rephrase these as \quotes{user-to-user} and \quotes{user-to-business} privacy relationships. Our work falls under the second category, where the data flow is between a user and a service provider.

\subsubsection{Privacy Awareness and Behaviors.}
While many web and mobile service providers offer privacy settings, users tend to accept default privacy settings without change. For example, a 2005 paper by Gross and Acquisti analyzed Facebook profiles of over 4000 university students and found that only 0.06\% restricted their profile information visibility~\cite{gross05}. While subsequent studies in social media privacy showed increased use of social media privacy settings over time~\cite{stutzman12}, researchers found that use of user-to-business privacy settings remains low. For example, in 2021 Farke et al.'s study on Google's My Activity found that only 35\% of the participants used Google's My Activity page~\cite{farke21}. Malkin et al.'s study on Google and Amazon smart speaker users found that only a minority made use of privacy settings~\cite{malkin19}. For example, of those who knew about the deletion feature, 67.9\% never used it~\cite{malkin19}. Similarly, Lau et al.'s study on smart speaker users found that most users did not use the privacy settings of their devices~\cite{lau18}. 

Several studies found that users lack awareness about privacy settings and the data that service providers collect. Farke et al. found that only 12\% indicated they were \quotes{extremely aware} of Google's My Activity, while most of them were \quotes{moderately aware} (35\%) and \quotes{somewhat aware} (28\%)~\cite{farke21}. Malkin et al. found that less than half (48.3\%) of smart speaker users knew that their audio recordings were saved permanently in their smart speaker devices, while 41.4\% incorrectly believed their recordings were saved temporarily~\cite{malkin19}. Moreover, of the 44\% of participants who were aware of the review feature, nearly half (45\%) were not aware that they can use it to delete recordings in their smart speaker devices~\cite{malkin19}. Zheng et al. reported that participants had a blurred distinction between device manufacturers and external parties that collect and save data in IoT devices~\cite{zheng18}. 

Researchers have suggested that users' lack of awareness and use of privacy settings is in part due to discoverability issues. For example, Habib et al. found that many users found difficulties in finding the \quotes{Ad Preferences}~\cite{habib22} and the opt-out~\cite{habib20} settings. Chen et al. examined 100,000 applications' privacy settings and classified 36.29\% of privacy settings as \quotes{hidden,} with 82.16\% of them permissive by default~\cite{chen19}. In addition, lack of knowledge about privacy settings was mentioned among the reasons for UK users' failures to change the mobile phone manufacturer's default settings~\cite{ramokapane19}. Likewise, users' overtrust in device manufacturers to protect their privacy sometimes leads them to assume that there are no additional steps required from them~\cite{zheng18}, and losing trust in device manufacturers can lead users to believe that using privacy settings is useless~\cite{lau18}. Finally, Willis provided a theoretical framework of the mechanisms that make default settings \quotes{sticky} (i.e., unlikely to be changed by users), which inspired initial a priori themes in our qualitative analysis of the reasons not to change defaults~\cite{willis14}.

\subsubsection{Privacy Preferences, Attitudes, and Concerns.}
Several studies investigated users' preferences about data sharing practices, such as which uses of data by device manufacturers users found acceptable. Malkin et al. and Zheng et al. studies on smart home devices showed that users found it acceptable to use their data for improving the service provided to them~\cite{malkin19,zheng18}. On the other hand, multiple studies reported that participants found use of their data for ad purposes to be unacceptable~\cite{malkin19,lau18,habib22}, especially if the ads are from a third party~\cite{malkin19,lau18}. Zheng et al. found divided opinions among participants regarding using their data for ads, and suggested that users' acceptance of using their data is benefit-driven~\cite{zheng18}. Another aspect of users' preferences that has been investigated in prior work is the data retention period. Malkin et al. found that smart speaker users preferred shorter retention periods over longer ones~\cite{malkin19}. Khan et al. found similar preferences among cloud storage users~\cite{khan18}. 

Several studies reported participants' attitudes when they viewed the service provider's actual behavior with respect to their data. Farke et al. reported that 33\% were surprised, and 35\% indicated that the amount of data was more than they anticipated~\cite{farke21}. Malkin et al. reported that many participants were surprised that their voice interactions with smart speakers were permanently saved, and that they could review them~\cite{malkin19}. Similarly, multiple studies on mobile apps reported that most participants were surprised about the amount~\cite{jung12}, frequency, and destinations~\cite{balebako13}, of the data that their mobile apps were collecting. 

In terms of concerns, Farke et al. found that exposure to Google's My Activity reduced participants who had privacy concerns from 52\% to 47\%~\cite{farke21}. However, several other studies on US participants in other contexts such as smart speakers and smart home devices reported fewer privacy concerns. Lau et al., Zheng et al., and Zeng et al. studies on smart home devices reported that participants had few privacy concerns~\cite{lau18,zheng18,zeng17}. Reasons for low concerns include trusting the manufacturer, not feeling targeted, incomplete understanding of the privacy risks associated with the device, and trading privacy for functionality or convenience~\cite{lau18,zheng18,zeng17}. Malkin et al. reported only 28.3\% of participants had privacy concerns about their smart speakers~\cite{malkin19}. However, Malkin et al. also noted that users raised higher concerns in specific contexts such as when recordings contain children's voices or when audio recordings are used for ads~\cite{malkin19}, offering contextual integrity as an explanation~\cite{nissenbaum04,malkin19}.

\subsection{Saudi Society and Privacy} \label{sec:saudi}
Saudi Arabia is a developing country in southwestern Asia that was established in 1932~\cite{wikipedia21}, and has one of the youngest populations in the world. As of 2019, 49\% of the population was below 30 years old~\cite{gas19_pop}. It has around 17\% of the world’s proven oil reserves, and is the second-largest member of the Organization of the Petroleum Exporting Countries (OPEC)~\cite{opec20}. Saudi Arabia's economy and young population resulted in rapid development and technology adoption. The Internet penetration rate in Saudi Arabia reached 89\% in 2019~\cite{gas19_internet}. It is reported that over 80\% of the Saudi population are social media users~\cite{kemp22,blogger22}. However, as of this writing, Internet users in Saudi Arabia have no legal protections for their personal data. While a draft of the Personal Data Protection Law (PDPL) was issued on September 2021~\cite{gazette21}, full enforcement of the law was postponed until September 2023~\cite{gazette23}. 

Despite high Internet usage among Saudis, very little online privacy research has been conducted with this population, and none that we are aware of investigated user-to-business privacy issues. Rashidi et al. surveyed 626 WhatsApp users in Saudi Arabia to understand their privacy behaviors and attitudes~\cite{rashidi16}. They found that Saudis were aware of WhatsApp privacy settings and used them, especially to limit the visibility of the \quotes{last seen} setting that shows when they were last seen active~\cite{rashidi16}. Another study by Alsagri and Alaboodi analyzed the privacy awareness and attitudes of 455 Snapchat users from Saudi Arabia~\cite{alsagri15}. They found that over 70\% of participants were concerned about \quotes{misuse or abuse of their personal information}~\cite{alsagri15}.

\section{Methodology} \label{sec:method}
We interviewed 30 Google personal account holders in Saudi Arabia about their awareness, attitudes, preferences, and concerns regarding the activity data that Google saves about them, as well as any steps they take to control Google’s collection or use of this data. In this section, we describe our recruitment procedure, interview procedure, analysis methods, and limitations. We obtained formal ethical approval for the study from King Abdulaziz University's research ethics committee. We obtained participants' explicit consent before taking part in the screening survey and interview. Participants were informed that the interview would be recorded and that the results of this study will be published without identifying participants. Moreover, they were informed that they can quit the study at any stage. 

\subsection{Recruitment} \label{sec:rec}
Our targeted sample was those who identified themselves as: living in Saudi Arabia (nationals or residents), aged 18 or older, using a free personal (not a business) Google Gmail account with \quotes{@gmail.com} extension regularly for at least a month. The rationale behind limiting the study to free personal @gmail.com accounts is that business accounts may be configured by the organization's Information Technology (IT) department and not by the users themselves. We aim to understand users' privacy decisions, not those of the organization. Moreover, paid accounts may have different default privacy settings than free accounts. In the invitation, we also stated that participants must use a personal computer that contains a web browser and Zoom (an online conference software~\cite{zoom22}). We required desktop or laptop computers because the Google account mobile interface may be different from the computer interface. Our study targeted non-security-expert users as we are more interested in understanding the baseline non-security-expert user perception and behavior. Moreover, asking security experts about their awareness and usage of privacy settings in an interview setting can make the results more prone to social desirability bias. We screened for university degree(s) and current job unrelated to cybersecurity by asking for degree and job generally, without mentioning security. 

To recruit participants, we sent an invitation for the screening survey written in Arabic to King Abdulaziz University's mailing lists, which included academic and administrative staff and students, who come from diverse backgrounds. King Abdulaziz University is one of the largest public universities in Saudi Arabia~\cite{kau23}. In addition, the first author sent invitations to  personal and professional networks, mainly through WhatsApp groups~\cite{whatsapp22}, which are widely used in Saudi Arabia, and posted invitations on Twitter~\cite{twitter22}, Facebook~\cite{facebook22}, and LinkedIn~\cite{linkedin22}. We  encouraged the recipients to share the invitation with their personal and professional networks, especially networks that encompass diverse demographics. Finally, at the end of each interview, we encouraged participants to invite others to take part. Participation was voluntary. We used a screening survey to check that participants met the inclusion criteria before  inviting them to the interview. The study's advertised title was: \quotes{Users' Experiences on Cloud Services.} We avoided using any words related to \quotes{security} and \quotes{privacy} in the study's title, invitation letter, screening survey, and during the interview, until the very last sections of the interview, to avoid biased answers that do not reflect users' actual privacy behaviors and decisions~\cite{sotirakopoulos11}. We informed participants that they will be invited to an interview that will include a user experiment on their Google Gmail account, where the interviewer will guide them step by step. We informed them that the experiment does not require any technical background or preparation. We also informed them that we will not request access to their devices, ask them to download any file, request any changes to their email, or request their password or any private data. We received 91 completed screening survey responses, we disqualified 27 that did not meet the study criteria. From the 64 qualified screening responses, we invited a demographically diverse sample of participants. Upon scheduling the interview, an email containing a Zoom link for the online interview was sent to the participants. Interview invitations were sent in batches. If an invited participant did not schedule an interview within 24--48 hours, canceled, or scheduled one but did not show up, we invited a replacement participant. This process was repeated until we reached the desired number of completed interviews (30).

\subsection{Interview} \label{sec:interview}
The interview was designed in a semi-structured format. We prepared a set of multiple-choice, open-ended, and task-based questions. We based some of our questions off of questions from Malkin et al.'s survey~\cite{malkin19}. The interview included a user experiment section in which we asked participants to log in to their Gmail accounts using their computers and web browsers. Then, we asked them questions about their settings, awareness, and usage of Google's Activity Controls. We also asked them about their sentiments after viewing their saved activity data, their attitudes regarding the most private data for them, their attitudes regarding Google's practices with their data, and finally their privacy concerns. We asked follow-up questions where needed. Interviews were conducted online using Zoom, an online conference software~\cite{zoom22}, from August 5 to 28, 2021. The interviewer shared their screen with participants via Zoom and displayed each question on the screen as they asked it (except for follow-up questions). In a few exceptional cases, participants voluntarily shared their screens if they needed help, e.g. to locate the right setting. However, we resumed the interview with the interviewer's shared screen once help was provided. Interviews were voice and screen recorded (the interviewer's screen only, with few exceptions to provide help). They were conducted in Arabic language by the first author, who is a native Arabic speaker and proficient in English. Each interview lasted for 33 minutes on average (mean: 33, minimum: 22, and maximum: 45 minutes).

\subsection{Data Analysis}\label{sec:analysis}
After we finished all interviews, recordings were manually transcribed by the first author and two paid undergraduate students. All transcripts were  reviewed by the first author. The interview questions (see~\autorefappendix{app:questions}) and the participants' qualitative answers included in our analysis were manually translated from Arabic to English by the first author. Multiple-choice questions were quantitatively analyzed using descriptive statistics. Open-ended questions included in our analysis were qualitatively analyzed using template analysis, a style of thematic analysis~\cite{king22,brooks14}, by a single trained researcher. The coder met with the co-author over multiple sessions to discuss the results and refine coding and themes. While multiple coders can confirm consistent interpretations, a single coder is deemed acceptable in Human-Computer Interaction (HCI) research~\cite{mcdonald19}. In template analysis, for each open-ended question, we started with \quotes{a priori themes} (the template), which were mainly inspired by our prior knowledge of the topic and the literature. We then read the data multiple times, and adjusted (added, updated, or removed) the themes accordingly. Finally, we mapped participants' answers to the suitable themes and sub-themes (coding the template)~\cite{king22}. 

\subsection{Limitations} \label{sec:limitations}
First, our interviews were conducted in Arabic to reach a wide range of Saudi Arabian participants and to enable rich expression using the participants' native language. As a result, the interview questions in~\autorefappendix{app:questions} and the direct quotes from our participants in this paper are the closest translations of the original questions and quotes from Arabic to English. We tried to accurately translate the interview questions and the participants’ answers, including the answers' imperfections. However, \quotes{lost in translation} expressions may have occurred. Second, our participants' demographics are biased toward those located in the Western region of Saudi Arabia (93\%), and slightly towards females (63\%). The regional bias is very likely due to the fact that the main author was based in the Western region of Saudi Arabia (Jeddah city) and that one of the main survey distribution channels was a large public university at the Western region (King Abdulaziz University). However, the Western region (a.k.a. Makkah Region) is a very large and diverse region, containing several cities and towns with a population of \num{9033491} according to the latest population estimates of Makkah region in mid 2019~\cite{gas19_pop}. King Abdulaziz University is the second largest public university in Saudi Arabia, comprising students coming from different backgrounds and regions in Saudi Arabia. As of 2019, it had over \num{165490} students, \num{7527} faculty staff, and \num{6739} administrative staff~\cite{gas19_uni}. Third, some participants' answers might be prone to recall bias and social desirability. To minimize recall bias, our interview utilized task-based questions, where the participants opened their actual account page and answered based on their accounts' actual settings. To minimize social desirability, we informed the participants at the beginning of the interview that there is no right or wrong answer, we are only interested in their choices and opinions, and we do not have a connection to Google. Moreover, we did not use words related to \quotes{security} and \quotes{privacy} until the very last sections of the interview, to avoid answers biased towards security and privacy behavior. Fourth, during our study, Google made some updates (detailed in~\autoref{sec:google}), which included updating a few terms. They changed the term \quotes{Data \& Personalization} to \quotes{Data \& Privacy,} and the term \quotes{Activity Controls} to \quotes{History Settings.} These updated terms were reported by four participants during the study, and we proceeded using the new terms (verbally), requiring early mention of the word \quotes{privacy} with these participants. Interface updates by service providers during field studies are not uncommon, and were reported by similar studies~\cite{malkin19,wang14,almuhimedi15}. Fifth, while we offer some comparisons between our study and those conducted in the US, \textbf{this is not a replication of a prior study.} Furthermore, this study, and most of the studies we compare it to, are qualitative studies. Thus, \textbf{we only observe trends that appear similar or different across cultures, but cannot make direct quantitative comparisons.} Finally, while we mark participants who reported having technical background with a superscript star (P\#\textsuperscript{*}) to make better sense of the data and quotes, analyzing different perspectives based on technical literacy is beyond our study's scope. 

\section{Results} \label{sec:results}
In what follows, we summarize our results. Participants are denoted by P, followed by the participant number (P\#). Participants with technical backgrounds are denoted by a superscript star (P\#\textsuperscript{*}). While qualitative analysis is meant to distill themes and not to quantify, we provide the number of participants whose responses were mapped to a particular theme to provide a better sense of our data. See~\autoref{fig:diagram} in~\autorefappendix{app:diagram} for an illustration summarizing our qualitative analysis themes.

\subsection{Participants} \label{sec:responses} 
After conducting three pilot interviews (not included in our results), we completed and analyzed 30 interviews. All of our participants were living in Saudi Arabia. Participants included 19 (63\%)  females and 11 (37\%)  males, ranging in age between 18 and 54 years old. Eight participants (27\%) reported having technical backgrounds (i.e. having a university degree or working in Computer Science (CS), Information Systems (IS), Information Technology (IT), or Computer Engineering (CE) areas). None of them reported a degree or work in the cybersecurity field (confirmed via a question at the end of the interview). 
~\autoref{tab:demo} provides detailed demographics of the participants.
\begin{table}[!t]
	\centering
	\caption{Participants' general demographics.}
	\label{tab:demo}
    \resizebox{0.35\textwidth}{!}{%
	\begin{tabular}{ll@{\hspace{5pt}}r}
		\toprule
		\multicolumn{3}{c}{\textit{N = 30}} \\
		\hline
		Nationality		& 	\#	&	\%		\\ 
		\hline
		\quad Saudi 	& 25 & (83\%) \\
		\quad Other    	& 5  & (17\%)  \\
		\hline
		Gender			 & 	\#	&	\%	\\
		\hline
		\quad Female 	& 19 & (63\%) \\
		\quad Male    	& 11 & (37\%)  \\
		\hline
		Residence region of Saudi Arabia			 & 	\#	&	\%	\\
		\hline
		\quad Western 	 						 & 28 & (93\%) \\
		\quad Eastern	 						 & 0 & (0\%) \\
		\quad Central    						 & 2 & (7\%) \\
		\quad Northern	 						 & 0 & (0\%) \\
		\quad Southern	 						 & 0 & (0\%) \\
		\quad Other (*required: please specify)	 & 0 & (0\%) \\
		\hline 
		Age				&  \#	& \% \\
		\hline
		\quad 18 to 24		& 10 & (33\%) \\
		\quad 25 to 34		& 10 & (33\%) \\
		\quad 35 to 44		& 5 & (17\%) \\
		\quad 45 to 54 		& 5 & (17\%) \\
		\quad 55 to 64 		& 0 & (0\%) \\
		\quad 65 or above	& 0 & (0\%) \\
		\hline 
		Highest degree completed &  \#	& \% \\
		\hline
		\quad Doctorate								&  2   &   (7\%)      \\
		\quad Master's								&  9   &   (30\%)      \\
		\quad Bachelor								&  11  &   (37\%)      \\
		\quad High school							&  8   &   (27\%)      \\
		\quad Intermediate school					&  0   &   (0\%)      \\
		\quad Elementary school						&  0   &   (0\%)      \\
		\quad Other (*required: please specify)		&  0   &   (0\%)      \\
		\hline
		Have CS/IS/IT/CE background?				&  \#	& \% \\
		\hline
		\quad Yes							& 8 & (27\%)\\
		\quad No							& 22 & (73\%) \\
		\hline
		Employment status							&  \#	& \% \\
		\hline
		\quad Student								& 11  & (37\%) \\
		\quad Full-time employee					& 11  & (37\%) \\
		\quad Part-time employee					& 0  & (0\%) \\
		\quad Self-employed or business owner		& 2  & (7\%) \\
		\quad Full-time house wife/husband			& 1  & (3\%) \\
		\quad Unemployed \& looking for a job		& 5  & (17\%) \\
		\quad Unemployed \& not looking for a job   & 0  & (0\%) \\
		\quad Unable to work						& 0  & (0\%) \\
		\quad Retired								& 0  & (0\%) \\
		\quad Other (*required: please specify)		& 0  & (0\%) \\
        \hline
		Google account age		& 	\#	&	\%		\\ 
		\hline
		\quad At least since a month  				& 0 & (0\%) \\
		\quad At least since 3 months					& 1 & (3\%) \\
		\quad At least since 6 months					& 2 & (7\%)  \\
		\quad At least since a year					& 2 & (7\%) \\
		\quad At least since 2 years					& 4 & (13\%) \\
		\quad At least since 3 years					& 4 & (13\%) \\
		\quad At least since 5 years					& 4 & (13\%) \\
		\quad At least since more than 5 years 		& 13 & (43\%) \\
		\quad Other (*required: please specify)		& 0  & (0\%) \\
  		\hline
		Browser type		& 	\#	&	\%		\\ 
		\hline
		\quad Google Chrome 		& 22 & (73\%) \\ 
		\quad Firefox				& 2	 & (7\%) \\
		\quad Brave				& 0	 & (0\%) \\
		\quad MS Edge				& 1  &  (3\%) \\
		\quad Safari				& 5  & (17\%) \\
		\quad Opera 				& 0  & (0\%) \\
		\quad Other (*required: please specify)	& 0  & (0\%) \\
		\bottomrule
	\end{tabular}
 } 
\end{table}

\subsection{Awareness of Google's Activity Controls} \label{sec:awareness}
\subsubsection{Users' Basic Activity Controls Settings vs. Defaults.} \label{sec:basic_settings} 
\begin{figure}[t!]
	\begin{center}
	\includegraphics[width=0.5\textwidth]{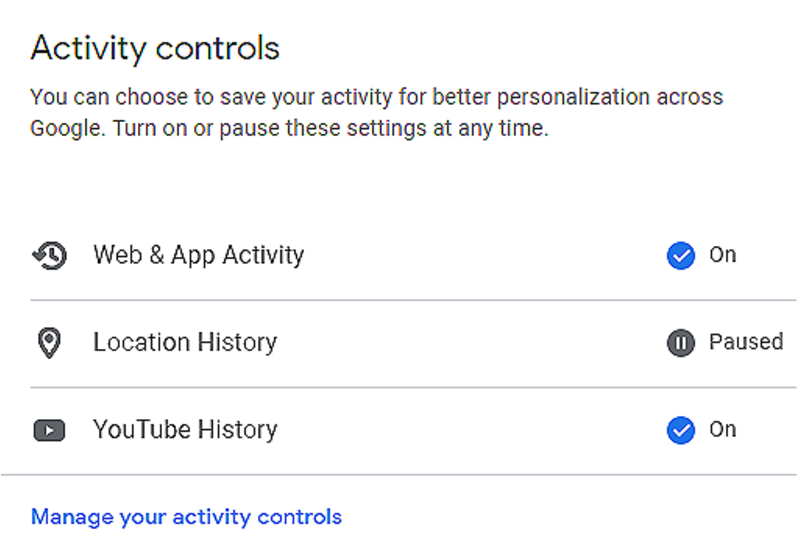}
    \end{center}
	\caption{Google's Activity Controls section.}
	\label{fig:activity_controls_area}
\end{figure}

Before we asked the participants about their awareness of Google's Activity Controls, we asked them to report their current (at the time of the study) basic Activity Controls settings. We guided them to the Activity Controls area in their Google accounts (see~\autoref{fig:activity_controls_area} for an illustration). Then, we asked them to report their Activity Controls basic settings (either \quotes{on} or \quotes{paused}) regarding saving three types of data: Web \& App Activity, Location History, and YouTube History (see Q. 1 in~\autorefappendix{app:questions}). We observe that for both the Web \& App Activity and YouTube History, most participants (28/30) kept the permissive default (\quotes{on}). On the other hand, for the Location History, where the default setting is restrictive (\quotes{paused}), we observe a smaller majority of participants (20/30) kept the default. See~\autoref{tab:basic_settings} for a full breakdown of users' basic Activity Controls settings.
\begin{table}[!t] 
	\centering
	\caption{Participants' basic Activity Controls settings compare to Google's default settings (Q.1 in~\autorefappendix{app:questions}).}
	\label{tab:basic_settings}
	\resizebox{0.6\columnwidth}{!}{%
		\begin{tabular}{lll@{\hspace{5pt}}rl@{\hspace{5pt}}r}
			\toprule
			\multicolumn{6}{c}{\textit{N = 30}} \\
			\hline
			\multirow{2}{*}{Data Type} & \multirow{2}{*}{Default Settings} & \multicolumn{4}{c}{Participants' Settings} \\
			\cline{3-6}
			&  & \multicolumn{2}{c}{On} & \multicolumn{2}{c}{Paused} \\
			\hline					   
			Web \& App Activity	& On 		& 28/30 & (93\%) & 2/30 & (7\%) \\
			Location History		& Paused 	& 10/30 & (33\%) & 20/30 & (67\%) \\
			YouTube History		& On 		& 28/30 & (93\%) & 2/30 & (7\%) \\
			\bottomrule
		\end{tabular}
	} 
\end{table}

\subsubsection{Awareness Level.} \label{sec:awareness_about_ac}
To identify participants' awareness level about Google's Activity Controls, we first briefly described the Activity Controls to the participants as follows: \quotes{allows you to control, such as turn on or pause saving your activities such as (Web \& App Activity, Location History, or YouTube History).} In addition, we shared Google's definitions of the following terms with participants~\cite{google21}:
\begin{inparaenum}[1)]
	\item Web \& App Activity: \quotes{Saves your activity on Google sites and apps, including associated info like location.}
	\item Location History: \quotes{Saves where you go with your devices, even when you aren't using a specific Google service.}
	\item YouTube History: \quotes{Saves the YouTube videos you watch and the things you search for on YouTube.}
\end{inparaenum} We then asked the participants if they were aware of Google's Activity Controls (see Q. 2 in~\autorefappendix{app:questions}). The majority of participants stated they had some level of awareness about Google's Activity Controls (24/30). However, of those 24 participants, many had only a vague awareness (i.e. answered \quotes{I expected there is such a thing} (9), or \quotes{I heard about such a thing} (2)), and only 13 answered \quotes{yes.}


\subsubsection{How Users Became Aware of the Activity Controls?} \label{sec:priv_how_became_aware}
For the 24 participants who expressed some level of prior awareness about Google's Activity Controls, we asked them how they knew about, heard of, or expected, the existence of the Activity Controls (see Q. 3 in~\autorefappendix{app:questions}). We then qualitatively analyzed their answers. From the 23 participants who provided answers (one participant did not remember), we identify seven themes: experiences with or expectations about applications and systems,  exploration and search, observed behavior of applications and systems,  chance, social channels,  consent prompts, and account sign-up or device setup. In what follows, we elaborate on each theme. 
\par\textbf{Experiences with or expectations about applications and systems:} 7/23 participants referred to their experiences with, or expectations about, applications and systems, when discussing how they became aware of Google's Activity Controls. 
Our participants mentioned experiences with applications and systems from Google and other service providers. In addition, three participants mentioned their experience with their browsers' history interface. P5 was one of two participants who mentioned that they were aware that they can delete their history from YouTube: \quotes{I know the YouTube History, but the Location History, and that there is a window called the Activity Controls that combines them all, I did not know honestly.} P25 mentioned his experience \quotes{on websites and controlling them.} On the other hand, P3 and P6 mentioned mere expectations of Google's Activity Controls, as P3 said: \quotes{I was expecting that we can control the activity ... because not everyone likes to have their activities recorded and kept.} 
\par\textbf{Exploration and search:} 5/23 participants mentioned becoming aware of Google's Activity Controls through exploration and search. For example,  P8\textsuperscript{*} said: \quotes{By experimentation ... from time to time I open the settings and I look and read them every now and then ....}
\par \textbf{Observed behavior of applications and systems:} 3/23 participants observed personalized application behavior that led them to expect that controls would exist. P11 described this experience: \quotes{while using the device, I see that the things I save, search about, and so on, get saved somewhere, but where? I do not know.} P20 said: \quotes{once you open YouTube you always find the videos that you recently watched ... and I do not know how exactly.}
\par \textbf{Chance:} 3/23 participants mentioned coming across Google's Activity Controls by chance (accidentally). For example, P27 said: \quotes{I wanted to change my account's password ... and found these topics on the activity and YouTube History ....} P14\textsuperscript{*} described a similar situation in addition to his awareness through the account's sign-up, he added: \quotes{some of them, when I want to edit the theme, or if for example, I want to add the payment and subscription and these things, I go to Data \& Personalization, I wonder what is this ....}
	
\par \textbf{Social channels:} 3/23 participants mentioned becoming aware of Google's Activity Controls through social channels. P7 and P17 mentioned WhatsApp~\cite{whatsapp22}. P17 said: \quotes{from the messages I receive on WhatsApp ... They were talking about that Google Map[s] knows your location, where you are, and so on, and how to modify it, how to turn it off ....} P26 mentioned Twitter: \quotes{I read on Twitter that there are things we are supposed to change to reduce the annoying ads.} 
\par \textbf{Consent prompts:} 
2/23 participants mentioned that consent dialogues made them expect that Google's Activity Controls would exist. 
For example, P23\textsuperscript{*} described browsers' prompts for users' permission to share the user's location with websites: \quotes{sometimes when I search on Google it tells me: do you want to turn on the location or not?} P9 mentioned consent prompts but was not specific on the type of application: \quotes{They usually write these notes: do you agree? Click on agree and you can change it from settings later.} 
\par \textbf{Account sign-up or device setup:} 
only 2/23 participants mentioned they knew about Google's Activity Controls during Google's account sign-up or device setup. 
For example, P14\textsuperscript{*} described his experience creating his account: \quotes{When I first created it ... one of the things was to make the Activity Controls or to skip them, if I am not mistaken.} P15 attributed her awareness to Google's phone (Pixel~\cite{pixel22}) setup: \quotes{I used to have a Google mobile ...  When I first turned on the device, I activated my account. I used to control these things from the mobile.} 


\subsection{Usage of Google's Activity Controls} \label{sec:usage}
\subsubsection{Usage Level.} \label{sec:priv_usage} 
We asked the participants if they have ever used Google's Activity Controls (see Q. 4 in~\autorefappendix{app:questions}). Despite the fact that 24 participants had expressed some level of awareness of the controls, only 11/30 said they had used them, and one did not remember. 

\subsubsection{Reasons for Using the Activity Controls.} \label{sec:priv_usage_reasons}
We asked the 11 participants who said they used Google's Activity Controls why they used the controls, and what changes they made (see Q. 5 and Q. 6 in~\autorefappendix{app:questions}). We then qualitatively analyzed the answers of both questions together. We identify four types of reasons to use Google's Activity Controls:  turning off data saving, turning on data saving, temporarily switching between data saving settings, and  reviewing or deleting saved data. In what follows, we elaborate on each theme. 

\par \textbf{Turning off data saving:}
7/11 participants mentioned using Google's Activity Controls to \quotes{turn off} (\quotes{pause} in Google's Activity Controls terms) data saving for one or more types of activity data. Six of these participants mentioned turning off the Location History, with some mentioning privacy concerns. We noted that the Location History was turned off by default in Google accounts at the time of the study. We don't know whether the Location History was turned on by default at some point in the past and these participants turned it off, or whether those participants were confused because the Activity Controls interface does not indicate which setting is the default. Only one participant mentioned turning off  the Web \& App Activity or the YouTube History, and three participants mentioned turning off data saving using general terms, such as turning off \quotes{tracking} and \quotes{history.} P10\textsuperscript{*} changed her settings to avoid targeted ads and explained how ads can be embarrassing in remote working settings: \quotes{Sometimes, or mostly, in our remote work, we share the screen ...  if you search about something, it will show immediately ... it shows to the co-workers with you. Sometimes it is something personal, not something you share with people.}
\par \textbf{Turning on data saving:}
In contrast to turning off data saving, only 2/11 participants mentioned using the Activity Controls to turn on data saving for one or more types of data. P4 mentioned using the Activity Controls to \quotes{turn on the YouTube History,} mainly to review his YouTube history as he described: \quotes{For example, there is something I want to recall, see it again, review it.} Similar to what we observed in the previous section, YouTube History was turned on by default in Google accounts at the time of the study. Thus, we don't know whether the participant turned it on (for example, if it was turned off by default at some point in the past), or the participant was confused because the Activity Controls interface does not indicate whether a user has selected a setting other than the default. P33 mentioned \quotes{I personalized the ads,} and we counted it as turning on data saving, although we note that Google's Ad Personalization settings are not actually part of the Activity Controls.

\par \textbf{Temporarily switching between data saving settings:}
3/11 participants mentioned temporarily switching between data saving settings. P8\textsuperscript{*} mentioned switching the Web \& App Activity setting between on and off depending on the situation. For example, if she is using multiple devices, she turns the Web \& App Activity on, to track the activities on those multiple devices, and turns it off if she is using a single device. P31 mentioned switching settings to reduce performance overhead: \quotes{the tracking and these things, in general, all take resources.} This is a misconception as Google saves this data in the cloud and the amount of activity data stored does not impact the performance of the user's device or account.  
\par \textbf{Reviewing or deleting saved data:}
One participant (P24) mentioned that she used Google's Activity Controls to review and delete data: \quotes{there are things that I search about which I prefer to delete, while I keep [other] things to return to them later.}
\subsubsection{Reasons for Keeping Default Settings.} \label{sec:priv_not_usage_reasons}
For the 18 participants who said they have not used Google's Activity Controls, we asked them about the reasons for keeping the defaults (see Q. 7 in~\autorefappendix{app:questions}). We qualitatively analyzed their answers and identify four main themes: high cost of change, perceived convenience, the \quotes{I have nothing to hide} attitude, and  low benefit of change. In what follows, we elaborate on each theme. 
\par \textbf{High cost of change:}
8/18 participants mentioned reasons relating to the high cost of change. We identify five types of costs as follows: \begin{inparaenum}[1)] \item \textit{cost of knowledge:} four participants said they did not change the default settings because they did not know about the Activity Controls, such as P28: \quotes{I do not know about it honestly.} \item \textit{cost of attention or care:} two participants mentioned that they did not pay attention to, or care about, the Activity Controls. For example P3 stated: \quotes{I do not pay attention to the settings and these things.}
\item \textit{cost of time:} refers to the time needed to explore, learn about, or change Google's Activity Controls. Two participants mentioned lack of time as a reason for sticking to the defaults. For example P13\textsuperscript{*} explained: \quotes{I do not go to the account and its settings unless I faced a problem ... I do not have enough time, to be honest.}
\item \textit{cost of cognitive memory:} refers to the cognitive memory required to remember to change the Activity Controls. This was mentioned by one pilot participant\footnote{Although this point was mentioned only by a pilot participant, we believe it is a reason worth surfacing.}. While some users may overcome the knowledge, attention or care, and time costs described above, they forget to change the Activity Controls. 
\item \textit{fear of \quotes{messing things up}:} this was mentioned by P17 who said: \quotes{Leave it as it is as long as the email is working ... The fear is to make something then this messes up the email or something like that.}
\end{inparaenum}

\par \textbf{Perceived convenience:}
7/18 participants mentioned convenience as a reason for sticking to the default Activity Controls settings. Perceived convenience is expressed in several phrases such as \quotes{It did not bother me,} \quotes{I did not face a problem,} \quotes{I find the History useful to me,} and \quotes{it is very convenient, I find the things that I frequently search about,} as stated by P5, P13\textsuperscript{*}, P19, and P20, consequently. 
\par \textbf{The \quotes{I have nothing to hide} attitude:}
4/19 participants expressed that they have nothing to hide because they have not done anything wrong or embarrassing that they would need to hide, or because their data is not of \quotes{high importance} (P16\textsuperscript{*}). 
P9 stated directly: \quotes{I have nothing to hide ....} 
\par \textbf{Low benefit of change:}
3/18 participants stated that they didn't think they would get much, if any, privacy or user experience benefits from changing their settings. Two participants expressed resignation because Google might save their information anyway or they might be tracked by other parties. 
%
%
P18 said: \quotes{because I know that even if I turn off the settings, there might be tracing and tracking for the websites I visit.} Similarly, P9 believes that giant service providers such as Google \quotes{if they want to save something legally or illegally, they will know how.} P6 believes that using the Activity Controls \quotes{will not make an impact or difference on usage.}


\subsection{Attitudes and Preference Toward Google's Data Practices and Activity Controls} \label{sec:attitudes}
\subsubsection{Users' Advanced Activity Controls Settings vs. Defaults.} \label{sec:advanced_settings}
\begin{figure}[!t]
	\centering
	\includegraphics[width=0.5\columnwidth]{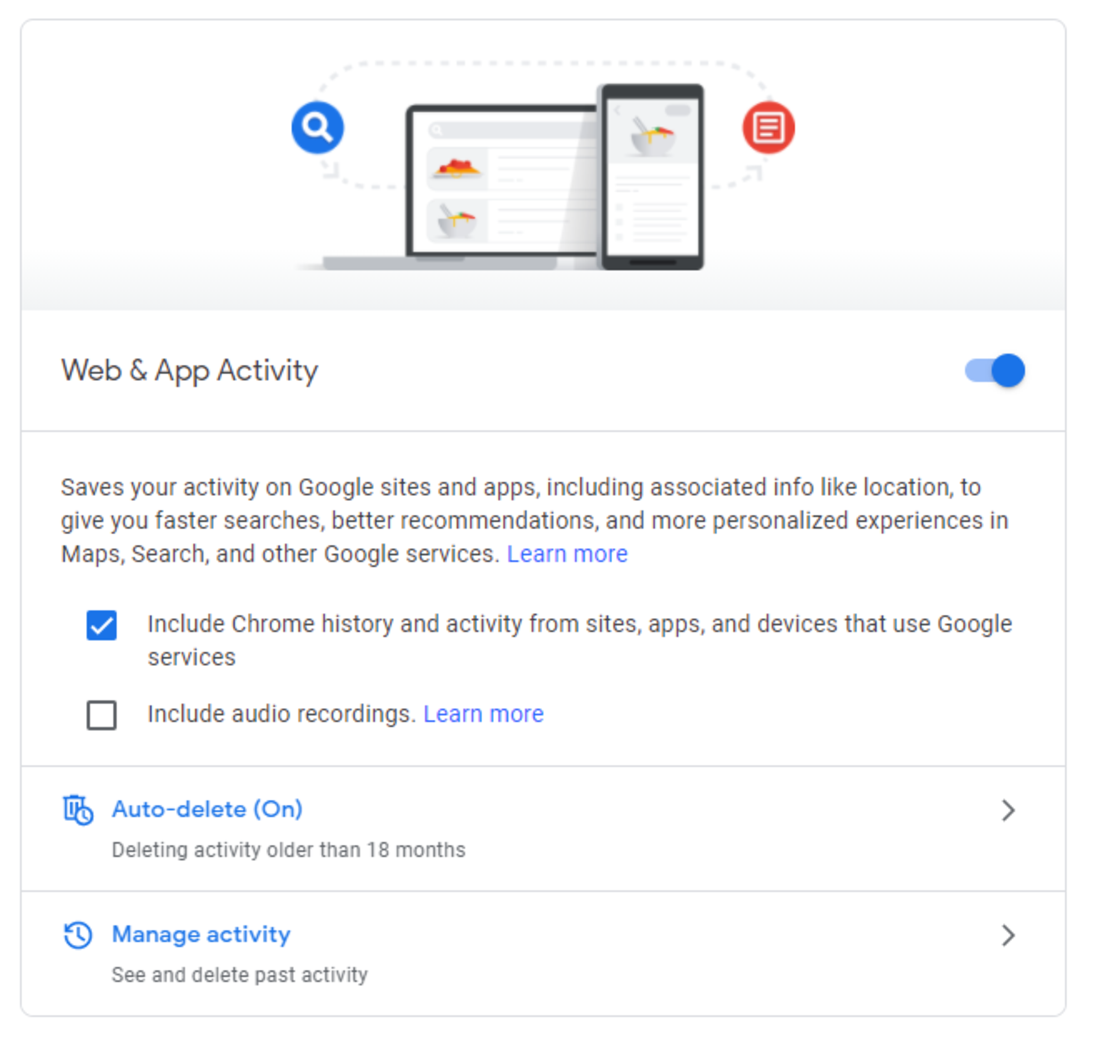}
	\caption{The advanced settings of the Web \& App Activity.}
	\label{fig:web_advanced_settings}
\end{figure}

Before we asked the participants about their attitudes toward Google's data saving practices, we asked them to report their advanced Activity Controls settings (see~\autoref{fig:web_advanced_settings} for an example of the Web \& App Activity advanced settings). To avoid interview fatigue, we proceeded to the advanced Activity Controls settings only on one of the Activity Controls that the participant reported had a status of \quotes{on} (see~\autoref{sec:basic_settings}). If the participant reported the Web \& App Activity was \quotes{on,} we proceeded to the Web \& App Activity for the advanced settings (28/30). Otherwise, if the participant reported the Web \& App Activity was \quotes{paused} and the YouTube History was \quotes{on,} we proceeded to the YouTube History for the advanced settings (1/30). If the participant reported all of the three Activity Controls were \quotes{paused,} we did not proceed with that participant to the advanced settings (1/30). We then asked the participants to report their advanced options (see Q. 8 in~\autorefappendix{app:questions}) and Auto-delete settings (see Q. 9 and Q. 10 in ~\autorefappendix{app:questions}). See~\autoref{tab:advanced_settings} and~\autoref{tab:autodelete_settings} for a full breakdown of those results, compared to the defaults.

\begin{table*}[!t]
	\centering
	\caption{Participants' Web \& App Activity and YouTube History Activity Controls advanced options (see Q. 8 in~\autorefappendix{app:questions}). \textit{N} represents the participants who reported one or more activity data types in their Activity Controls settings are \quotes{on.} We omitted the Location History settings as the Location History did not have any advanced options (check-boxes), and we did not proceed to the Location History advanced settings with any participant.}
	\label{tab:advanced_settings}
    \resizebox{0.8\textwidth}{!}{%
	\begin{tabularx}{\textwidth}{lX|c|cc|cc}
		\toprule
		\multicolumn{7}{c}{\textit{N = 29}} \\
		\hline
		\multirow{2}{*}{Data Type} & \multirow{2}{*}{Options} & \multirow{2}{*}{Default Settings} & \multicolumn{4}{c}{Participants' Settings} \\
		\cline{4-7}
		& 						&  						 & \multicolumn{2}{c}{checked} & \multicolumn{2}{c}{unchecked}   \\ 
		\cline{4-7}
		& 						&  						 & \# & \% & \# & \%  \\
		\midrule
		\multirow{2}{*}{Web \& App Activity} 
		& \tabitem Include Chrome history and activity from sites, apps, and devices that use Google services  & checked	 & 24
		/28 & (86\%)  & 4/28 & (14\%) \\
		\cline{2-7}
		& \tabitem Include audio recordings 																   	& unchecked  	 & 0/28  & (0\%) & 28/28 & (100\%)  \\
		
		\midrule
		\multirow{2}{*}{YouTube History}   	 
		& \tabitem Include the YouTube videos you watch 													    & checked 	& 1/1 & (100\%) & 0/1 & (0\%) \\
		\cline{2-7}
		& \tabitem Include your searches on YouTube 															& checked	& 1/1 & (100\%) & 0/1 & (0\%) \\
		\bottomrule
	\end{tabularx}
    } 
\end{table*}

\begin{table*}[!t]
	\centering
	\caption{Participants' Auto-delete and retention period settings in Google's Activity Controls. See Q.9 and Q.10 in~\autorefappendix{app:questions}. Note that the retention period is applicable only if the Auto-delete setting is \quotes{on.} \textit{N} represents the participants who reported one or more activity data types in their Activity Controls settings are \quotes{on.} We omitted the Location History settings as we did not need to proceed to it with any of our participants.}
	\label{tab:autodelete_settings}
	\resizebox{0.8\textwidth}{!}{%
		\begin{tabularx}{\textwidth}{lll|lr|lr|lr|lr|lr} 
			\toprule
			\multicolumn{13}{c}{\textit{N = 30}} \\
			\hline
			\multirow{3}{*}{Data Type}   & \multicolumn{2}{c}{Default Settings}  & \multicolumn{9}{c}{Participants' Settings}\\
			\cline{2-3}\cline{4-13}
			& \multirow{2}{*}{Status} &	\multirow{2}{*}{Delete Every}  & \multicolumn{4}{c}{Status}  & \multicolumn{6}{c}{Delete Every} \\
			
			\cline{4-13}			
			& 			&              & \multicolumn{2}{c}{On}  & \multicolumn{2}{c}{Off}  & \multicolumn{2}{c}{3 mo.}& \multicolumn{2}{c}{18 mo.}& \multicolumn{2}{c}{36 mo.} \\ 
			
			\midrule
			& 			&              & \# & \% & \# & \% & \# & \% & \# & \% &  \# & \% \\
			\cline{4-13}
			
			Web \& App Activity & \quotes{On} 			&  18 mo.      & 6/28 & (21\%) & 22/28 & (79\%)	 	& 4/6 & (67\%)  & 2/6 & (33\%) & 0/6 & (0\%)  \\

			\midrule
			
			YouTube History	 & \quotes{On} 			&  36 mo.     & 1/1  & (100\%) &  0/1  & (0\%) 	& 1/1  & (100\%) & 0/1  & (0\%) & 0/1 & (0\%) \\
\hline
            & Total& & 7/29 & (24\%) & 22/29 & (76\%) & 5/7 & (71\%) & 2/7 & (29\%) & 0/7 & (0\%) \\
			\bottomrule   
		\end{tabularx}
	}
\end{table*}

\subsubsection{Attitudes Towards Data Saving.} \label{sec:sentiments}
Following the protocol described in the previous section to select one Activity Controls data type to proceed to the attitude question on, we guided the 29 participants (28 on the Web \& App Activity and 1 on YouTube History) to the \quotes{Manage data} link (see~\autoref{fig:web_advanced} for an example in the Web \& App Activity), which allows them to \quotes{See and delete past activity.} Then, we asked them if they found records of any of their previous searches (see Q. 11 in~\autorefappendix{app:questions}). All  29 participants found saved activity data in their accounts. After they viewed some of their saved data, we asked them to describe their feeling (see Q. 12 in~\autorefappendix{app:questions}). We qualitatively analyzed their answers and classified their feelings toward the saved data into three themes:  negative, neutral, and positive. 
We elaborate on each theme.    
\par \textbf{Negative Sentiment:}
13/29 participants expressed negative sentiments after they viewed their saved data in Google's Activity Controls. This led several of them to remark that going forward they would change their behavior: two said they plan to be more aware of what they do on the Internet, and four said they will turn off data saving or delete their data. We identify several motivations behind the negative sentiments. \begin{inparaenum}[1)] 
\item \textit{feeling watched:} seven participants mentioned that they felt watched. For example, P27 said: \quotes{I feel tracked} and P26 said: \quotes{it is easy to surveil me} in their sentiments. 
\item \textit{lack of, or incomplete, knowledge about the service provider's data practices:} six participants mentioned words that express lack of, or incomplete, knowledge about Google's data practices, such as \quotes{surprised,} \quotes{shock,} \quotes{scary,} \quotes{if I knew that,} and \quotes{did not expect that.} Most of those participants have incomplete knowledge about aspects of how the data are being saved, such as the breadth, depth, retention period, and linkability of the saved data. 
\item \textit{data breadth:} five participants mentioned surprise at seeing the amount of data saved. The phrase \quotes{everything is recorded} was used by multiple participants (P3, P23\textsuperscript{*}). For example, P3 said: \quotes{everything is recorded, I mean everything is stored and recorded in the History. So this matter is a bit scary.} Also, P23\textsuperscript{*} said: \quotes{I do not know about this, and that everything is recorded about me.}
\item \textit{linkability:} two participants were not aware that their activity data are linked to their accounts, as opposed to only being saved locally in the browser. For example, P25 said: \quotes{I know that in the browser, but in Google [account], this is the first time I know that this information appears.}
\item \textit{data depth:} which refers to the amount of details, was mentioned by one participant: \quotes{even the devices that I ... use, where I am, and the location.} (P23\textsuperscript{*}). 
\item \textit{retention period:} one participant (P5) mentioned concern about how long the data was saved: \quotes{old things, I mean very old History ... I want to turn off ... I want to delete ....}

\end{inparaenum}
\par \textbf{Neutral sentiment:} 10/29 participants expressed neutral sentiments after they viewed their saved data in Google's Activity Controls. We identify three motivations behind the neutral sentiments.  
\begin{inparaenum}[1)]
\item \textit{awareness or expectations of data saving:} five participants remarked that, simply viewing their activity data was not very surprising to them because they were already aware that Google saves their data. Multiple participants mentioned the word \quotes{normal.} For example, P21\textsuperscript{*} said: \quotes{Normal ... as a technical person because I know that Google stores these things.} Similarly, P8\textsuperscript{*} expected that her data are saved through \quotes{personalized advertising and such things, it is clear that it is using my search.}
\item \textit{the \quotes{I have nothing to hide} attitude:} three participants expressed the feeling of having nothing to hide. The statements \quotes{Nothing is worthy,} \quotes{my search is very general,} and \quotes{The search I have is purely for work,} were mentioned by P10\textsuperscript{*}, P16\textsuperscript{*}, and P17, respectively.
\item \textit{resignation:} some participants felt powerless towards Google, and believe there are no privacy settings that can stop giant companies such as Google from collecting and saving their data. This reason motivated two participants to express neutral sentiments. P14\textsuperscript{*} described the activity data saving by Google as \quotes{inevitable devil} and noted \quotes{any computer you use, your data will go, even if the company claimed secrecy [i.e., privacy], you cannot ensure that. That's why I did not bother myself by turning it off.}
\end{inparaenum}
\par \textbf{Positive sentiment:} Only 6/29 participants expressed positive sentiments after they viewed their saved activity data in Google's Activity Controls. All of those participants referred to the usefulness of being able to retrieve or review their stored data when needed. It is worth noting that none of those participants who expressed positive sentiments have a technical background. We identify two motivations behind the positive sentiments. 
\begin{inparaenum}[1)] 
\item \textit{retrieve records:} four participants mentioned being able to retrieve lost or forgotten records. For example, P31 said: \quotes{I am happy that I find it, because sometimes I remember that there is something I searched about, so I go here and find it immediately.}
\item \textit{review records:} three participants mentioned reviewing activity records such as searches or YouTube videos. For example, P4 said: \quotes{from time to time I like to see what did I do previously, what did I search about ... and watch it again.} This theme captures participants' desire to review content a second time rather than retrieve forgotten records.
\end{inparaenum}

\subsubsection{Informing Users about Data Saving.} \label{sec:suggestions}
We asked the 29 participants who reported one or more basic Activity Controls were “on” (see~\autoref{sec:basic_settings}) whether they were aware that Google saves this data about them (see Q. 13 in~\autorefappendix{app:questions}). We observe a pattern similar to the awareness of the Activity Controls. Here we also find an overall high, but vague, awareness of data saving (i.e., many answered \quotes{I expected ...} or \quotes{I heard about ...} that Google saved activity data about them). We find 11/29 answered \quotes{yes,} 6/29 answered \quotes{no,} while 8/29 answered \quotes{I expected that, but I am not certain,} and 4/29 answered \quotes{I heard about that, but I am not certain.} For the 18 participants who were not fully aware (i.e. their answer was not \quotes{yes}) that their data are being saved by Google, we asked them what would have informed them about the saving of their activity data (see Q. 14 in~\autorefappendix{app:questions}). We qualitatively analyzed their answers and identify three main themes: transparency, awareness, and nothing can be done. 

\par  \textbf{Transparency:}
10/18 participants mentioned suggestions relating to improving the clarity of communication about Google's data practices and the choices offered to users.
\begin{inparaenum}[1)] \item \textit{Explicit consent:} five participants said that an agreement about data saving should be clearly stated during the account sign-up. However, several participants admitted that they might have skipped the agreement as P3 continued: \quotes{It might be there but we do not notice ....} However, there are reasons for ignoring the agreement, as P5 said: \quotes{because it was not very clear.} An additional participant (P16\textsuperscript{*}) advocated for privacy-protective default settings in which data would not be collected or saved unless the user explicitly consents: \quotes{similar to what happened recently in the iPhone when it asks you: do you want to share your activities with the app? this should be the right situation, that the default is no sharing, and the exception is for sharing.} Google asks for users’ consent to collect and use their data. However, their approach appears not to be transparent enough, as several participants did not recall seeing it.
\item \textit{Better presentation:} three participants suggested better presentations, including less text. P9 suggested using videos that show \quotes{how the saving method, usage method ....} P20 suggested using bullet points, while P11 suggested placing the Activity Controls \quotes{in an easier place, not in the account ....} The latter suggestion can explain why we observed multiple participants were aware of the YouTube data saving but not the Web \& App Activity and Location History. YouTube uses the term \quotes{Your data in YouTube} in the account's main menu on YouTube, not buried under multiple menus and using opaque terms such as Data \& Personalization and Activity Controls. 
 \item \textit{User engagement:} two participants suggested user engagement, such as a mandatory step-by-step interactive wizard to force users to configure their privacy settings as \quotes{it involves the user in this. It makes him do the thing by himself. But when they put text no one would read it ...} (P14\textsuperscript{*}). 
\end{inparaenum}

\par \textbf{Awareness:}
9/18 participants mentioned suggestions that relate to awareness, which can be classified into two main categories.
\begin{inparaenum}[1)]
\item \textit{Notifications or nudges:} eight participants mentioned using notifications or nudges to inform users that their activity data are being saved, remind them to delete their saved data, or notify them if sensitive data are being saved. Two participants mentioned in-browser notifications or nudges, such that \quotes{when someone visits a website it says it is going to store [the data]} (P13\textsuperscript{*}).
\item \textit{Social media:} one participant (P29) suggested that Google could use social media to inform users about privacy-related matters.  
\end{inparaenum}
\par  \textbf{Nothing can be done:}
One participant (P28) believes that nothing can be done and blamed herself: \quotes{Nothing else they can do. This is the only known way, that they tell you all the conditions for all the things they do, but our problem is that we do not read it.} 

\subsubsection{Knowledge and Usage of the Review, Delete, and Auto-delete Features.} \label{sec:awarenss_use_data_saving} 
We asked the 10  participants who answered \quotes{yes,} they know that Google saves data about them (see Q. 13 in~\autorefappendix{app:questions})\footnote{An additional participant said \quotes{yes} to this questions, but missed~\autoref{sec:awarenss_use_data_saving} 's questions due to an error.}, whether they knew that they can review the data that Google saves about them (see Q. 15 in~\autorefappendix{app:questions}). 5/10 answered \quotes{yes} (4 in the Web \& App Activity, and 1 in the YouTube History groups), while the remaining participants answered \quotes{no.} We asked them how often they review the data (see Q. 16 in~\autorefappendix{app:questions}). 3/5 answered \quotes{at least once every 6 months,} one answered \quotes{less than once a year,} and one answered \quotes{I never reviewed them.} We asked the same 10 participants who said they knew Google saves data about them whether they knew they can manually delete some or all the data that Google saves about them (see Q. 17 in~\autorefappendix{app:questions}), and whether they knew about the Auto-delete feature (see Q. 19 in~\autorefappendix{app:questions}). 7/10 said they knew they could manually delete data,  and only 3/10 said they knew about Auto delete. Of the 7 who said they knew about manual deletion, we asked them how often they manually delete their data (see Q. 18 in~\autorefappendix{app:questions}). We find 4/7 said they \quotes{never deleted them,} and 2/7 \quotes{do not remember.}

\subsubsection{Preferences Regarding Google's Default Data Retention Periods.} \label{sec:retention}
We asked all the 29 participants\footnote{This question was added after P1 interview was done. Thus, 29/30 participants were asked this question.} whether they find Google's default retention period for each type of activity data suitable (Web \& App Activity: 18 months, YouTube History: 36 months, Location History: 18 months. See Q. 20, Q. 22, and Q. 24 in~\autorefappendix{app:questions}). If they answered \quotes{no,} we asked them to propose a suitable period from their perspective (see Q. 21, Q. 23, and Q. 25 in~\autorefappendix{app:questions}). Our results suggest that the majority of participants find Google's default retention periods unsuitable for all three types of activity data. As the retention period increases, more participants rank it unsuitable. The majority of participants suggested lower retention periods. For the Web \& App activity, the top suggested periods are: 1, 3--6, and 12 months, mentioned by 6, 4, and 6 participants, respectively. For the YouTube History: 1, 6, and 12 months, mentioned by 5, 5, and 5 participants, respectively. Finally, for the Location History: 1 and 6 months, and 1 day, suggested by 7, 4, and 3 participants, respectively. Multiple participants mentioned that they prefer the Location History to never be saved (but we asked them to specify the minimum preferred period, assuming that the data saving is on). See~\autoref{tab:retention_rank} for a full breakdown of the results. We also asked all participants about their attitudes regarding the most private and sensitive data type for them. Participants ranked the Location History as the most sensitive data type with the highest level of privacy, while the YouTube History was the least. See~\autoref{tab:privacy_rank} and~\autoref{tab:sens_rank} for a full breakdown of the results.
\begin{table}[!t]
	\centering
	\caption{Participants' preferences regarding Google's default data retention periods.}
	\label{tab:retention_rank}
    \resizebox{0.6\textwidth}{!}{%
		\begin{tabular}{lcl@{\hspace{3pt}}rl@{\hspace{3pt}}r}
			
			\toprule
			\multicolumn{6}{c}{N = 29}\\
			\hline
			\multirow{2}{*}{Activity Type} 		& Default Retention Period & \multicolumn{4}{c}{Suitable} \\
			\cline{3-6}
			
			& 		(months)	& \multicolumn{2}{c}{Yes} & \multicolumn{2}{c}{No}  \\
			\hline					   
			Web \& App Activity	 	& 18 mo.		& 8 & (28\%)  & 21 & (72\%) 	  \\
			YouTube History			& 36 mo.		& 4 & (14\%)  & 25 & (86\%) 	  \\
			Location History		 	& 18 mo.		& 6 & (21\%)  & 23 & (79\%) 	   \\
			\bottomrule
		\end{tabular}
    } 
\end{table}
\begin{table}[!t]
	\centering
	\caption{Participants' ranking for the privacy degree for each type of activity data for them. See Q. 26 in~\autorefappendix{app:questions}.}
	\label{tab:privacy_rank}
	\resizebox{0.5\textwidth}{!}{%
		\begin{tabular}{llrlrlr}
			\toprule
            \multicolumn{7}{c}{\textit{N = 30}} \\
			\hline
			\multirow{2}{*}{Data Type} 		& \multicolumn{6}{c}{ Privacy Rank} \\
			\cline{2-7}
			& \multicolumn{2}{c}{High} & \multicolumn{2}{c}{Med} & \multicolumn{2}{c}{Low}\\
			\hline					   
			Web \& App activity	 	& 9 & (30\%) & 19 & (63\%) & 2 & (7\%) \\
			Location History		 	& 18 & (60\%) & 8 & (27\%) & 4 & (13\%)\\
			YouTube History			& 4 & (13\%) & 17 & (57\%) & 9 & (30\%) \\
			\bottomrule
		\end{tabular}
	 } 
\end{table}
\begin{table}[!t]
	\centering
	\caption{Participants' ranking for the sensitivity of each type of activity data for them. See Q. 27 in~\autorefappendix{app:questions}.}
	\label{tab:sens_rank}
	\resizebox{0.6\textwidth}{!}{%
		\begin{tabular}{ll@{\hspace{5pt}}rl@{\hspace{5pt}}rl@{\hspace{5pt}}rl@{\hspace{5pt}}r} 
			\toprule
            \multicolumn{9}{c}{\textit{N = 30}} \\
			\hline
			\multirow{3}{*}{Data Type} 				& \multicolumn{8}{c}{Sensitivity Rank} \\
			\cline{2-9}
			& \multicolumn{2}{l}{\makecell{All\\sensitive}} & \multicolumn{2}{l}{\makecell{Some\\sensitive}} & \multicolumn{2}{l}{\makecell{All\\non-Sensitive}} & \multicolumn{2}{c}{Other}\\
			
			\hline					   
			Web \& App Activity 		& 1  & (3\%)  & 25 & (83\%) & 4   & (13\%) & 0 & (0\%) \\
			Location History		 	& 19 & (63\%) & 4  & (13\%) & 7   & (23\%) & 0 & (0\%) \\
			YouTube History			& 2  & (7\%)  & 8  & (27\%) 	  & 20 	& (67\%)	& 0 & (0\%) \\
			\bottomrule
		\end{tabular}
	} 
\end{table}


\subsubsection{Preferences Regarding Acceptable Use of Activity Data.} \label{sec:accept_rank}
\begin{figure}[!h]
	\centering
	\includegraphics[width=\textwidth]{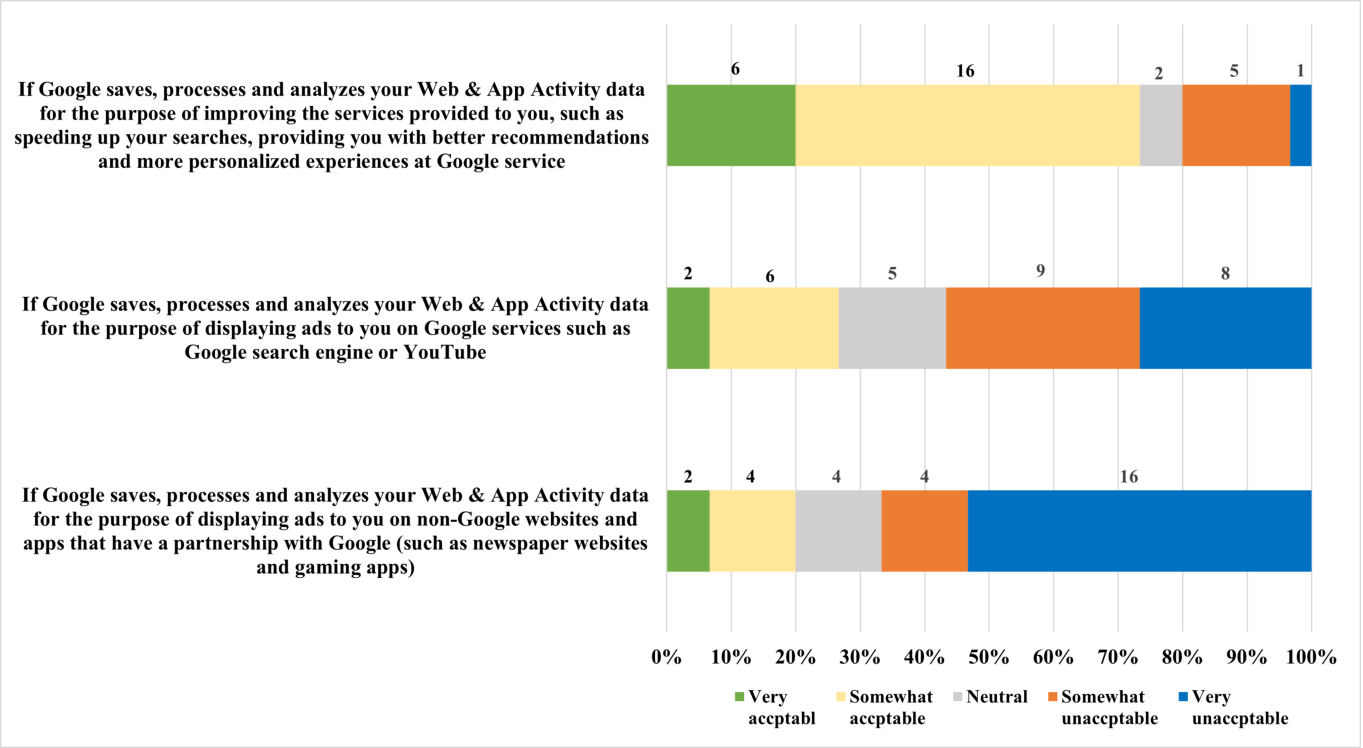}
	\caption{Participants' preferences regarding the acceptable use of their Web \& App Activity data by Google.}
	\label{fig:data_usage_rank}
\end{figure}

We presented all participants with three different scenarios of how Google may use their Web \& App Activity data, and asked them to rank how acceptable each scenario is for them (see Q. 28, Q. 29, and Q. 30 in~\autorefappendix{app:questions}). We limited these questions to the Web \& App Activity data to avoid interview fatigue. The Web \& App Activity is \quotes{on} by default and contains rich data about users. The results suggest that the majority of participants (22/30) would accept using their Web \& App Activity data to improve Google services. On the other hand, the majority would not accept using their data for ads displayed on Google services (17/30), or in non-Google websites and apps that partner with Google (20/30). See~\autoref{fig:data_usage_rank} for an illustration. 

\subsection{Privacy Concerns} \label{sec:concerns}
\subsubsection{Users' Privacy Concerns.} \label{sec:priv_concerns}
We asked all participants whether they previously had any privacy concerns while using Google services (see Q. 31 in~\autorefappendix{app:questions}). We find 13/30 participants answered \quotes{yes,} 15/30 answered \quotes{no,} and 2/30 were \quotes{neutral.} We did not ask the participants why did they have or have not any privacy concerns. However, some of them voluntarily elaborated their answers with the reasons. Out of the 9 participants who elaborated on why they did not have privacy concerns, we find 5 participants’ responses (P4, P9, P11, P20, P31) map to the \quotes{I have nothing to hide} category, because they  felt they did not have anything either sensitive, wrong, or important. P11 referred to her trust in Google as they are \quotes{an international company and aspire to make their [products] better,} while P20 added she did not have concerns because she is not a targeted person. For those who had privacy concerns, we asked them about these concerns (see Q. 32 in~\autorefappendix{app:questions}). 12/13 provided a comprehensible answer. We qualitatively analyzed their answers and 
classified  participants' privacy concerns into two categories: data abuse and data leak. In addition, we note that three participants referred to service providers such as Google as adversaries, two participants mentioned the words \quotes{hackers} and \quotes{hacking,} and one participant said she is more concerned about someone who knows her than Google.  

\par  \textbf{Data abuse:}
of the 13 participants who had privacy concerns, 10/13 participants mentioned concerns that relate to data abuse, noting five types of data abuse.
\begin{inparaenum}[1)]
\item \textit{Tracking or spying on users:} this is a common concern, mentioned by 7 participants. P22 explained: \quotes{I am concerned that Google knows a lot of data about me, knows my interests, knows where I go and where I come, what I search about} and described this situation as \quotes{somewhat an unpleasant idea.} P19 described her concern as: \quotes{someone can access my account and see what I am searching for.} The word \quotes{spying} was mentioned by one participant only. 
\item \textit{Using users' data for purposes they are unaware of:} two participants were concerned about using their data for purposes they are unaware of. For example, P8\textsuperscript{*} was concerned that her data are used \quotes{in a way that could harm me, or in a wrong way, or in a way that I do not accept.} 
\item \textit{Sharing users' data with other parties:} two participants mentioned concerns about users' data being shared with other parties. P8\textsuperscript{*} referred to the other party as \quotes{an entity that [the data] should not reach to,} while P14\textsuperscript{*} was concerned that \quotes{these big companies sell your data.}
\item \textit{User impersonation:} P7 was concerned about user impersonation where someone falsely claims to be a particular user.  
\item \textit{Intellectual property theft:} P7 described his concern as someone might take someone else's \quotes{ideas, and take what he is thinking about and what he is doing.}
\end{inparaenum}

\par \textbf{Data leak:}
5/13 participants were concerned that their data could be leaked, either accidentally, because of an attack on the service provider, or because of a data privacy breach. P29 was particularly concerned about a \quotes{photos} leak. Three participants linked data leaks with attacks on the service provider as P18 described: \quotes{we heard that even Google and even Apple were hacked and thousands or even hundreds of thousands of millions of personal data for their users were leaked.}


\subsubsection{Steps Users Took to Protect Their Privacy.} \label{sec:priv_concerns_steps}
We asked all participants if they have ever taken any steps to protect their privacy on the Internet while using Google services (see Q. 33 in~\autorefappendix{app:questions}). 18/30 answered \quotes{yes,} while 12/30 answered \quotes{no.} For those who answered \quotes{yes,} we asked them what are these steps (see Q. 34 in~\autorefappendix{app:questions}). We qualitatively analyzed their answers and identify three main themes: change privacy or security settings or control permissions, change browsing behavior, and use software assisting tools. In what follows, we elaborate on each theme. 

\par \textbf{Change privacy or security settings or control permissions:}
11/18 participants mentioned changing privacy or security settings or controlling permissions to protect their privacy. Six participants mentioned turning off the location data sharing in Google accounts and other settings. P13\textsuperscript{*} mentioned disallowing location sharing in most mobile apps except for government apps. P23\textsuperscript{*} mentioned disallowing location permission to websites when asked by the browser. P8\textsuperscript{*} mentioned turning on notifications for logins from new devices. P17 said he changed the password after his \quotes{email was hacked.} Finally, P29 mentioned enabling two-factor authentication. 

\par \textbf{Change browsing behavior:}
7/18 participants mentioned a wide variety of practices to change their Internet browsing behavior. P7 mentioned using a privacy-friendly search engine such as \quotes{DuckDuckGo}~\cite{duck22} as the main engine and using the Google search engine only if DuckDuckGo did not return results. P8\textsuperscript{*} would not use her account on a \quotes{public wi-fi.} P9 mentioned he \quotes{delete[s] the cookies from time to time.} P10\textsuperscript{*} mentioned deleting the YouTube history from the browser's history manually, P26 mentioned reviewing and deleting the account's YouTube History data for kids. P15 mentioned using a Virtual Private Network (VPN). P18 mentioned private browsing modes, such as Firefox's Private Browsing~\cite{mozilla22} and Google Chrome's Incognito mode~\cite{incognito22}. However, a VPN and private browsing cannot prevent a service provider such as Google from saving activity data inside the account.

\par \textbf{Use software assisting tools:}
2/18 participants mentioned using software assisting tools to protect their privacy online while using Google services. P7 mentioned using the \quotes{C Cleaner} software~\cite{ccleaner22} to clean the history, \quotes{trackers,} and \quotes{adware.} P7 also mentioned using the \quotes{Pocket}~\cite{pocket22} to \quotes{drop the links} he wants to save, while P9 mentioned using \quotes{Internet antivirus.}
\subsubsection{Future Steps Users Plan to Take to Protect their Privacy.} \label{sec:priv_concerns_future_steps}
We asked all participants whether they plan to take future steps to protect their privacy online when using Google services (see Q. 35 in~\autorefappendix{app:questions}). 17/30 answered \quotes{yes,} while 13/30 answered \quotes{no.} For those who answered \quotes{yes,} we asked them what these steps are (see Q. 36 in~\autorefappendix{app:questions}). We qualitatively analyzed their answers and identified two main themes: review or change privacy settings and use privacy-friendly systems or products.  

\par \textbf{Review or change privacy settings:}
out of the 17 participants who plan to take future steps to protect their privacy, 16/17 participants mentioned that they intend to review or change their privacy settings, in particular, Google's Activity Controls. Most participants want to change their settings to restrictive settings. For example, P23\textsuperscript{*} \quotes{will go to the settings and see how much data I am sharing,} P10\textsuperscript{*} will turn off the YouTube History and the Location History data saving in her Google account, and P14\textsuperscript{*} said that \quotes{the Web \& App Activity, once we finish I will change them.} Moreover, 8 participants mentioned they want to turn on the Auto-delete or reduce the data retention period. P12\textsuperscript{*} changed her retention period during the interview: \quotes{now while I am talking to you, I reduced it to 3 months for example.} This suggests that when users are empowered with knowledge, they care about their privacy and can take action to protect it. 

\par \textbf{Use privacy-friendly systems or products:}
one participant (P7) said, he \quotes{maybe moving from Google Chrome} to a more privacy-friendly browser and added: \quotes{I am currently using Brave~\cite{brave22} to check if it is better to use.} Moreover, he is considering returning to Linux OS as he thinks \quotes{Linux is more privacy-protecting than Windows.}

\section{Discussion and Cross-Cultural Perspective} \label{sec:discussion}
When it comes to user-to-business privacy, many service providers today adopt deceptive interfaces and permissive default settings. Our results showed that while most of our participants had some level of awareness about Google's Activity Controls, many of them were only vaguely aware, most of them had never used them, and many were shocked by the extent of the data saved about them. Our results also showed that many users care about their privacy and once informed about settings and data being saved, said they intend to take future steps to protect their privacy. However, it is unrealistic to expect users to learn how to identify deceptive interfaces and to assume that users will remember to change permissive default settings. Thus, our key recommendations from this research are mainly directed toward service providers, researchers, and policy makers. It has become clear that improved privacy setting interfaces are needed to inform users about service provider data practices and available choices. Research is needed to examine different techniques for when, how, and what to present in privacy settings to users. Service providers should deploy more usable privacy setting interfaces with privacy-friendly defaults and improved transparency. The challenge is to find the right balance between transparency and usability in privacy settings. Policy makers should call out service providers' deceptive techniques in privacy interfaces that result in  many users sharing more than intended. Where such practices are illegal under existing laws, regulators should go after offending service providers. 

Comparing our results with previous studies in the US, overall, we observe that our Saudi participants exhibit similar trends in privacy awareness, attitudes, preferences, concerns, and behaviors to what has been found in studies in the US. As detailed in the related work (\autoref{sec:related}), in terms of awareness, similar to US participants~\cite{farke21,malkin19,felt12,almuhimedi15,balebako13,zheng18}, Saudi participants showed \textit{vague awareness about data saving practices and privacy settings}. Similar to what has been found in multiple studies in the US~\cite{farke21,malkin19,balebako13,jung12}, when Saudi participants viewed the data Google saves about them, \textit{many were surprised}. Similar to US participants, Saudi participants found it \textit{acceptable} for a service provider to use their data for improving the services provided to them. On the other hand, they found it \textit{unacceptable} for a service provider to use their data for ads purposes~\cite{malkin19,lau18,habib22}, and to allow third parties to use their data~\cite{malkin19,lau18}. Furthermore, similar to US participants~\cite{malkin19,khan18}, Saudi participants tend to \textit{prefer shorter data retention periods}. In terms of behavior, similar to US participants~\cite{farke21,malkin19,lau18,habib22}, Saudi participants showed \textit{low usage of privacy settings}, and almost all \quotes{types} of behaviors that Saudi participants reported that they have taken previously, or plan to take in future, are common privacy behaviors reported by US participants~\cite{farke21,colnago23}. In terms of privacy concerns, similar to what was reported by US participants in the Google context~\cite{farke21}, many Saudi participants reported they previously had privacy concerns. However, the high contextuality of privacy concerns makes it difficult to compare across contexts, for example with studies done in the context of smart home devices~\cite{lau18, malkin19,zeng17,zheng18}. Since our study is not a replication of any of the US studies, we suggest further research to explore whether, and to what extent, cultural differences, and other underexplored contextual factors, such as the free (e.g. Gmail account) versus paid account or device (e.g. smart home device), the service provider's name and its relevant attributes (age, reputation, popularity), affect users' concerns. 

\section{Google Steps to Improve Privacy Settings} \label{sec:google}
\begin{figure} [!th]
	\centering
	\includegraphics[width=0.8\columnwidth]{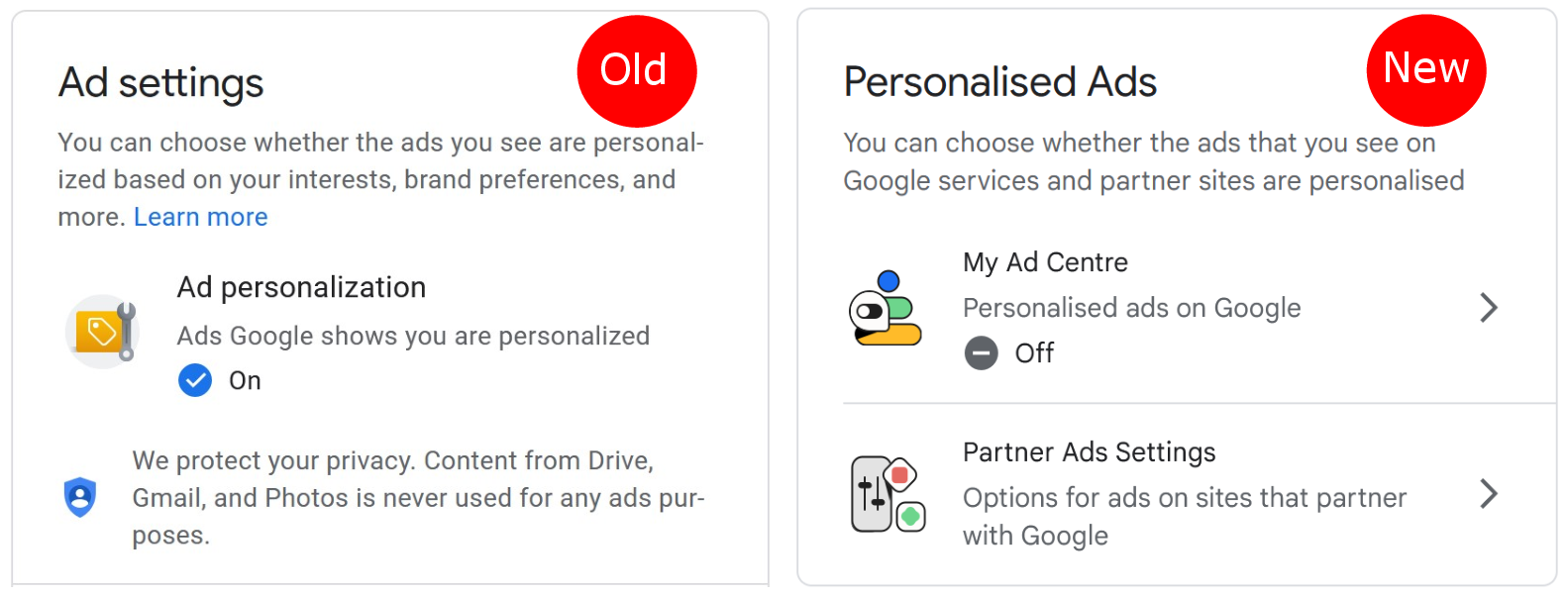}
	\caption{Google's updated ad settings}
	\label{fig:updated_ads_settings}
\end{figure}

\begin{figure} [!th]
	\centering
	\includegraphics[width=0.5\columnwidth]{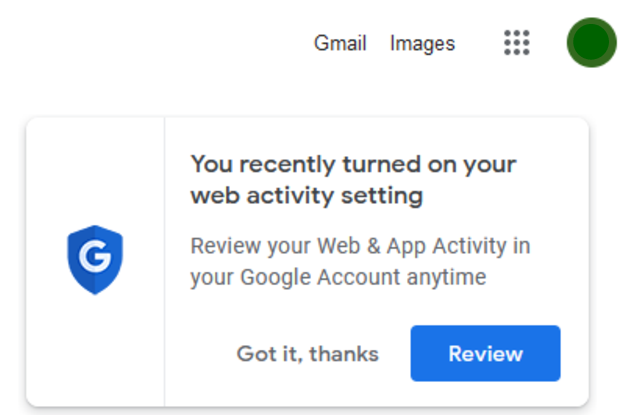}
	\caption{Google's Web \& App Activity nudge.}
	\label{fig:nudge}
\end{figure}

During and after our study, Google took some steps that are in line with our findings. First, they updated some terms, which support our theme of \quotes{transparency} through \quotes{better presentation} that would have informed unaware users about the data saving~(\autoref{sec:suggestions}). Namely, they updated the term Data \& Personalization to Data \& privacy, and the term Activity Controls to History Settings. Second, Google appears to have started launching nudges about the Activity Controls~(\autoref{fig:nudge}), which comes under the suggestion of \quotes{awareness} through \quotes{notifications or nudges} that would have informed unaware users about data saving~(\autoref{sec:suggestions}).~\autoref{fig:nudge} was first encountered by the main researcher a while after completing the study on Aug. 2022. However, we could not replicate the nudge even if we turned on the Activity Controls. Third, Google also updated their personalized ad settings with more customization~(\autoref{fig:updated_ads_settings}), which accommodates users' attitudes against using their data for Google partners' ads~(\autoref{fig:data_usage_rank}). Having said that, the paper provides other issues that may need Google's attention, such as shorter data retention periods. Moreover, despite Google updates, our results can guide other application providers.

\section{Conclusion} \label{sec:conclusion}
We interviewed 30 Google personal account holders in Saudi Arabia to understand their privacy perceptions and behaviors regarding the activity data that Google saves about them. Our study focused on Google's Activity Controls. Our results showed that although the majority of participants have some level of awareness about Google's Activity Controls and data practices, many have vague awareness about them, and the majority have not used the available controls. When the participants viewed their saved data in their accounts, many were surprised as they felt they were being watched, and they lack knowledge about Google's data practices. Many participants would accept Google's use of their data to improve the services provided to them, but would not accept using their data for ad purposes. Finally, we observed that our Saudi participants exhibited similar trends and patterns in privacy awareness, attitudes, preferences, concerns, and behaviors to what has been found in studies in the US. However, our study is not a replication of any of the US studies and further research is needed to directly compare US and Saudi participants. Our results emphasize the need for: \begin{inparaenum} \item improved techniques to inform users about privacy settings during account sign-up, to remind users about their settings, and to raise awareness about privacy settings; \item improved privacy setting interfaces to reduce the costs that deter many users from changing the settings; and \item further research to explore privacy concerns in non-Western cultures\end{inparaenum}.

\section{Acknowledgement} \label{sec:ack}
Eman Alashwali acknowledges the financial support of the Ibn Rushd Program at King Abdullah University of Science and Technology (KAUST). We thank all the participants for their time and valuable answers. We especially thank pilot participants for their valuable feedback. We thank Reema Bamakhramah and Sara Alashwali for their help in transcribing the interviews.

\bibliographystyle{ACM-Reference-Format}
\bibliography{ref}
\clearpage
\small
\appendix
\begin{landscape}
\section*{Appendices} \label{sec:appendices}
\section{Summary of our Qualitative Analysis Themes for Privacy Awareness, Attitudes, Concerns, and Behaviors} \label{app:diagram}
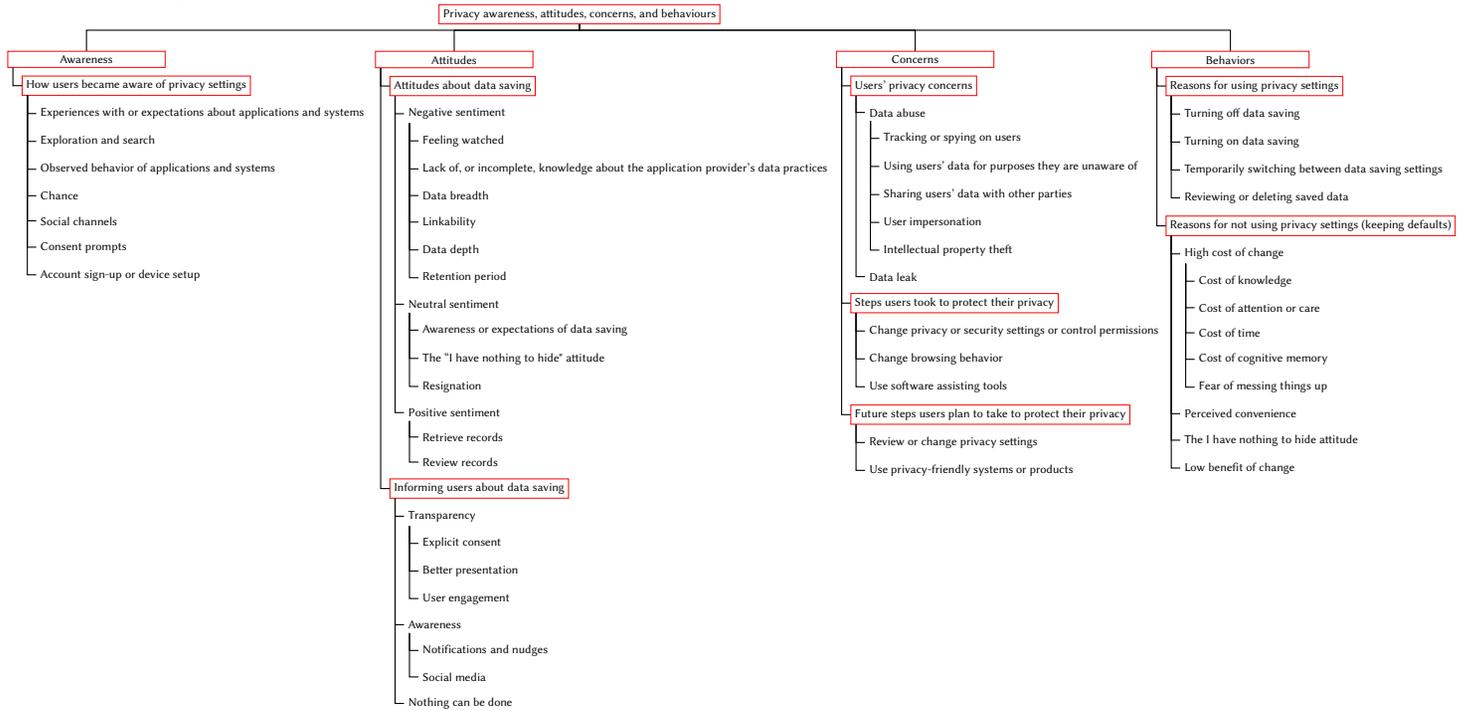
\begin{figure}[!th]
\centering
\caption{A diagram summarizing themes for privacy awareness, attitudes, concerns, and behaviors distilled from our qualitative data.}
\label{fig:diagram}
\begin{adjustbox}{width=1.28\textwidth}
\forestset{
	direction switch/.style={
		forked edges,
		for tree={
			edge+=thick, 
			font=\sffamily,
		},
		where level=1{minimum width=13em}{},
		where level<=2{draw=red}{},
		where level>=1{folder, grow'=0}{},
	},
}
	\newlength\gap
	\setlength\gap{10mm}

\begin{forest} 
		direction switch
		[Privacy awareness{,} attitudes{,} concerns{,} and behaviours 
			[Awareness
				[How users became aware of privacy settings
                    [Experiences with or expectations about applications and systems]
					[Exploration and search]
					[Observed behavior of applications and systems]
					[Chance]
					[Social channels]
					[Consent prompts]
					[Account sign-up or device setup]
				]
			] 
			[Attitudes
				[Attitudes about data saving
                    [Negative sentiment
                        [Feeling watched]
						[Lack of{,} or incomplete{,} knowledge about the application provider's data practices]
                        [Data breadth]
                        [Linkability]
                        [Data depth]
                        [Retention period]
					]
					[Neutral sentiment
						[Awareness or expectations of data saving]
						[The ``I have nothing to hide" attitude]
						[Resignation]
					]
                    [Positive sentiment
						[Retrieve records]
						[Review records]
					]
				]
				[Informing users about data saving
					[Transparency
						[Explicit consent]
						[Better presentation]
						[User engagement]
					] 
					[Awareness
						[Notifications and nudges]
						[Social media]
					]
					[Nothing can be done]
				]	
			] 
			[Concerns
				[Users' privacy concerns
					[Data abuse
						[Tracking or spying on users]
						[Using users' data for purposes they are unaware of]
						[Sharing users' data with other parties]
						[User impersonation]
						[Intellectual property theft]
					]
					[Data leak]
				]
				[Steps users took to protect their privacy
					[Change privacy or security settings or control permissions]
					[Change browsing behavior]
					[Use software assisting tools]
				]
				[Future steps users plan to take to protect their privacy
					[Review or change privacy settings]
					[Use privacy-friendly systems or products]
				]
			] 
			[Behaviors
				[Reasons for using privacy settings
					[Turning off data saving]
					[Turning on data saving]
					[Temporarily switching between data saving settings]
					[Reviewing or deleting saved data]
				]
				[Reasons for not using privacy settings (keeping defaults)
					[High cost of change
						[Cost of knowledge]
						[Cost of attention or care]
						[Cost of time]
						[Cost of cognitive memory]
						[Fear of messing things up]
					]
                    [Perceived convenience]
                    [The I have nothing to hide attitude]
					[Low benefit of change
					]
				]
			] 
		]
	\end{forest}
 
\end{adjustbox}
\end{figure}
\fillandplacepagenumber
\end{landscape}
\clearpage
\clearpage
\section{Interview Questions} \label{app:questions}
In what follows, we list the interview questions related to this paper. Text between square brackets is not shown to participants (added for clarification). There are more instructions, illustration figures, demographic, closing, and other questions. The full interview script is available upon request.
\subsection{Users' Experiment}
\begin{figure}[!th]
	\centering
	\includegraphics[width=0.5\columnwidth]{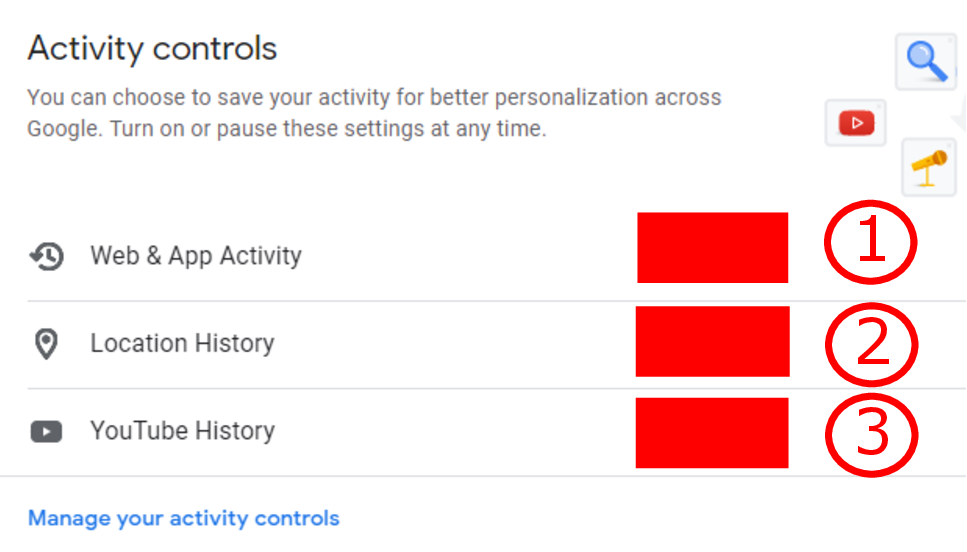}
	\caption{The Activity Controls basic settings. Participants were asked to report their accounts' current settings (either \quotes{on} or \quotes{paused}). We covered our settings in the figure to avoid bias or confusion.}
	\label{fig:activity_controls}
\end{figure}

[Questions about awareness of Google’s Activity Controls]
\begin{itemize}
    \item \textbf{Q. 1:} Which choice appears to you in the shaded areas numbers: 1, 2, and 3, in front of the following items: Web \& App Activity, Location History, and YouTube History? \par
    \textbf{Answers:} \begin{inparaitem}[$\circ$]\item \quotes{Stop,}\footnote{\label{note1}A fictitious answer was added as an answer validity check.} \item \quotes{On,} \item \quotes{Paused}\end{inparaitem} \par 
    \textbf{Related figures:}  \autoref{fig:activity_controls} 
    \vspace{4pt}
    \item \textbf{Q. 2:} Were you aware of Google's Activity Controls that enable you to control the activities that Google saves about you such as the Web \& App Activity, YouTube History, and Location History? (yes/no, elaborate if possible)? \par 	
    \textbf{Answers:} 
    \begin{inparaitem} [$\circ$] \item \quotes{Yes,} \item \quotes{No,} \item \quotes{I heard about such a thing,} \item \quotes{I expected there is such a thing} \end{inparaitem} \vspace{4pt}
	\item \textbf{Q. 3:} \textbf{[If Q. 2 answer is \quotes{yes,} \quotes{I heard about such a thing,} or \quotes{I expected there is such a thing}]} How did you know about, heard of, or expected the existence of Google's Activity Controls? \par 
	\textbf{Answers:} Open	
\end{itemize}
[Questions about usage of Google’s Activity Controls]
\begin{itemize}
	\item \textbf{Q. 4:} Have you ever used Google's Activity Controls? (yes/no, elaborate if possible). \par 
	\textbf{Answers:} \begin{inparaitem} [$\circ$] \item \quotes{Yes,} \item \quotes{No,} \item \quotes{I do not remember}\end{inparaitem}
 \vspace{4pt}
	\item \textbf{Q. 5:} \textbf{[If Q. 4 answer is \quotes{yes}]} Why did you use Google’s Activity Controls? \par 
	\textbf{Answers:} Open  
 \vspace{4pt}
	\item \textbf{Q. 6:} \textbf{[If Q. 4 answer is \quotes{yes}]} What changes did you make using Google's Activity Controls? \par 
	\textbf{Answers:} Open 
 \vspace{4pt}
	\item \textbf{Q. 7:} \textbf{[If Q. 4 answer is \quotes{no}]} If you have not used the Activity Controls before, it means you are keeping the default settings regarding the activities that Google saves about you, such as the Web \& App Activity and the YouTube History. What are your reasons for not using Google's Activity Controls and keeping the default settings? \par 
	\textbf{Answers:} Open 
\end{itemize}	

\begin{figure}[!th]
	\centering
	\includegraphics[width=0.5\columnwidth]{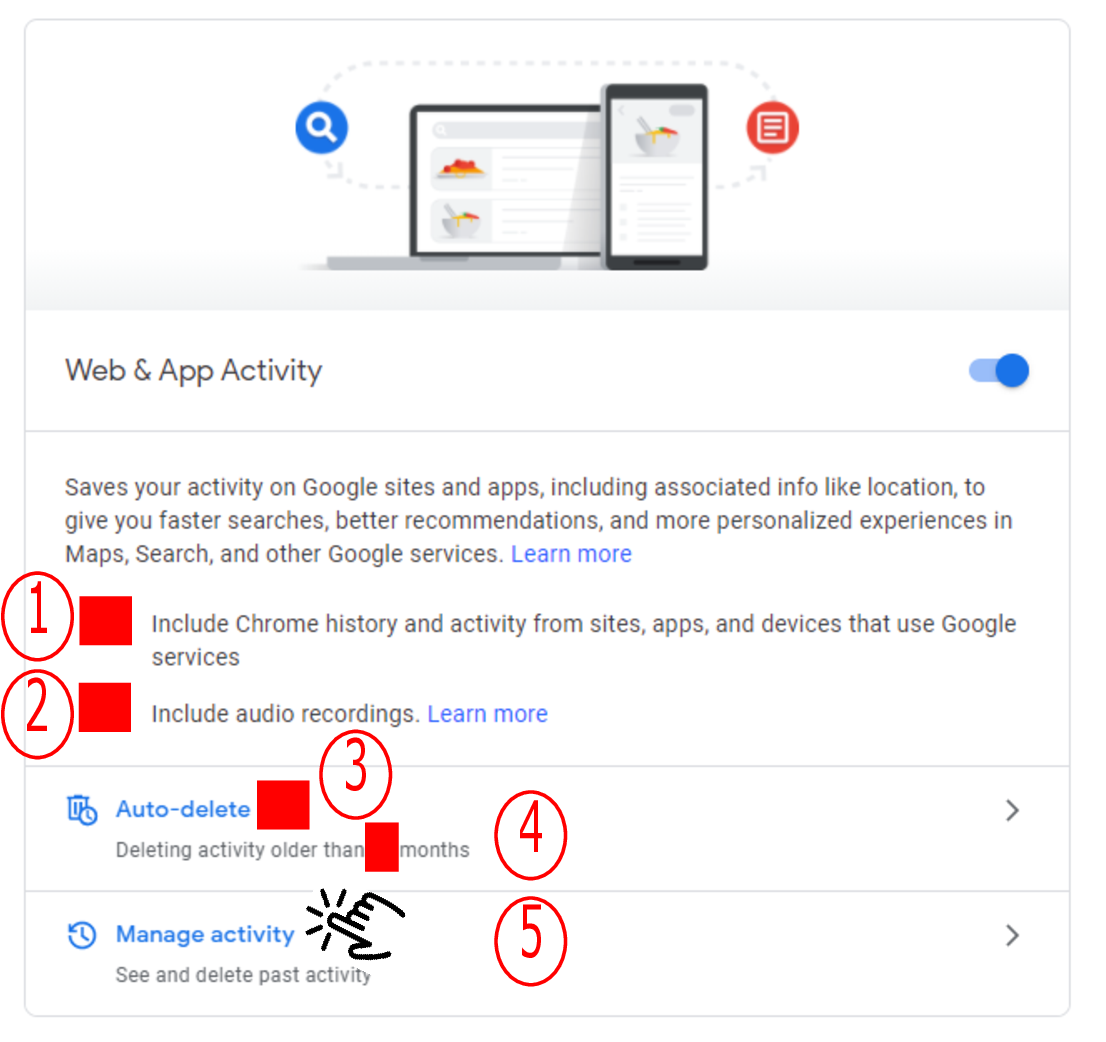}
	\caption{The advanced options of the Web \& App Activity. Participants were asked to report their settings (either checked \quotes{\cmark} or unchecked). We covered our settings in the figure to avoid bias or confusion.}
	\label{fig:web_advanced}
\end{figure}

\begin{figure} [!th]
	\centering
	\includegraphics[width=0.5\columnwidth]{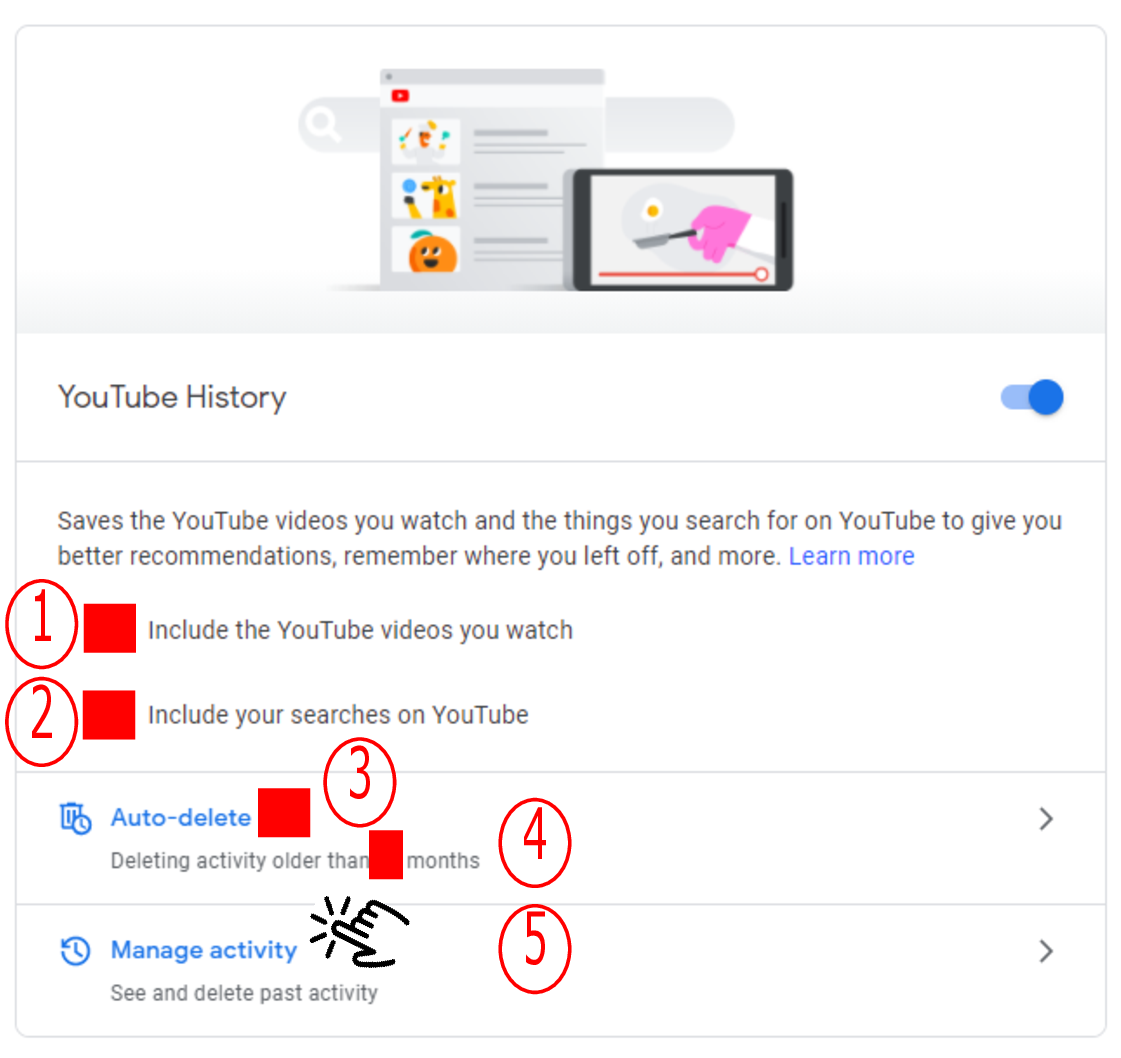}
	\caption{The advanced options of the YouTube History. Participants were asked to report their settings (either checked \quotes{\cmark} or unchecked). We covered our settings in the figure to avoid bias or confusion.}
	\label{fig:youtube_advanced}
\end{figure}

[Questions about preferences and attitudes towards Google's data practices and Activity Controls]
\begin{itemize}
    \item \textbf{Q. 8:} \textbf{[If Q. 1 answer is at least one \quotes{on}]} What are the checked choices in your [Web \& App Activity $|$ YouTube History $|$ Location History] \footnote{\label{note2}The participant was asked about either the Web \& App Activity, YouTube History, or Location history, depending on the participants' answer on the basic settings (Q. 1). The task description is in~\autoref{sec:advanced_settings}}? \par
    \textbf{Answers:} \par
    \textbf{[If we proceeded to the Web \& App Activity advanced settings, the following answers were shown]} \par 
    \begin{inparaitem}[$\square$] \item \quotes{Include Chrome history and activity from sites, apps, and devices that use Google services,} \item \quotes{Include audio recordings,} \item \quotes{None of the above,} \item  \quotes{Other (*required: please specify)}\end{inparaitem}  
     \vspace{2pt}
     
    \textbf{[If we proceeded to the YouTube History advanced settings, the following answers were shown]} \par 
    \begin{inparaitem}[$\square$] \item \quotes{Include the YouTube videos you watch,} \item \quotes{Include your searches on YouTube,} \item \quotes{None of the above,} \item  \quotes{Other (*required: please specify)}\end{inparaitem} \par 
    \textbf{Related figures:} \autoref{fig:web_advanced} and~\autoref{fig:youtube_advanced} 
     \vspace{2pt}
     
    \textbf{[If we proceeded to the Location History advanced settings]}\par 
    [No choices were presented as the Location History has no advanced options] 
    \vspace{4pt}
    \item \textbf{Q. 9:} \textbf{[If Q.1 answer is at least one \quotes{on}]} What is the status of the Auto-delete that appears to you\footref{note2}? \par
    \textbf{Answers:} \begin{inparaitem} [$\circ$] \item \quotes{On,} \item \quotes{Off,} \item \quotes{Not applicable,} \item \quotes{Stop,}\footref{note1} \item \quotes{Other (*required: please specify)} \end{inparaitem}\par 
    \textbf{Related figures:} \autoref{fig:web_advanced} and~\autoref{fig:youtube_advanced} 
    \vspace{4pt}
    
    \item \textbf{Q. 10:} \textbf{[If Q. 9 answer is \quotes{on}]} How long does Google save your data before automatic deletion?
    \footref{note2}? \par
    \textbf{Answers:} \begin{inparaitem} [$\circ$] \item \quotes{3 months,} \item \quotes{18 months,} \item \quotes{36 months,} \item \quotes{Other (*required: please specify)} \end{inparaitem} \par 
    \textbf{Related figures:} \autoref{fig:web_advanced} 
    \vspace{4pt}
    
	\item \textbf{Q. 11:} \textbf{[If Q.1 answer is at least one \quotes{on}]} Have you found [previous searches or so $|$ geographic locations] from the [Web \& App Activity $|$ YouTube History $|$ Location History]\footref{note2}? (yes/no, elaborate if possible). \par 
	\textbf{Answer:} \begin{inparaitem} [$\circ$] \item \quotes{Yes,} \item {no} \end{inparaitem}\par 
	\textbf{related Figures:} \autoref{fig:web_advanced} and~\autoref{fig:youtube_advanced}
 \vspace{4pt}

	\item \textbf{Q. 12:} \textbf{[If Q.11 answer is \quotes{yes}]} Describe your feeling after you found [Web \& App Activity $|$ YouTube History $|$ Location History]\footref{note2} data about you? \par 
	\textbf{Answers:} Open
 \vspace{4pt}
	
\item \textbf{Q. 13:} \textbf{[If Q.11 answer is \quotes{yes}]} Were you aware that Google saves the [Web \& App Activity $|$ YouTube History $|$ Location History]\footref{note2} data about you? \par

\textbf{Answer:} \begin{inparaitem}[$\circ$]  \item \quotes{Yes,} \item \quotes{No,} \item \quotes{I expected that, but I am not certain,} \item \quotes{I heard about that, but I am not certain}\end{inparaitem}
\vspace{4pt}

\item \textbf{Q. 14:}  \textbf{[If Q. 13 answer is \quotes{no}]} Do you have a suggestion if Google implemented, it would have informed you about the saving of your data? \par 
\textbf{Answers:} Open
\vspace{4pt}

\item \textbf{Q. 15:}  \textbf{[If Q. 13 answer is \quotes{yes}]} Did you know that you can review the [Web \& App Activity $|$ YouTube History $|$ Location History]\footref{note2} data that Google saves about you? \par
\textbf{Answers:} \begin{inparaitem}[$\circ$] \item \quotes{Yes,} \item \quotes{No}\end{inparaitem} 
\vspace{4pt}

\item \textbf{Q. 16:} \textbf{[If Q.15 answer is \quotes{yes}]} Approximately, how often do you review these data? \par 
\textbf{Answers:} \begin{inparaitem}[$\circ$] \item \quotes{At least once a day,} \item \quotes{At least once a week,} \quotes{At least once a month,} \quotes{At least once every 3 months,} \quotes{At least once every 6 months,} \quotes{At least once a year,} \quotes{Less than once a year,} \quotes{I never reviewed them,} \quotes{I do not remember}\end{inparaitem}
\vspace{4pt}

\item \textbf{Q. 17:} \textbf{[If Q.13 answer is \quotes{yes}]} Did you know that you can manually delete some or all of the [Web \& App Activity $|$ YouTube History $|$ Location History]\footref{note2} data that Google saves about you? 
\vspace{4pt}

\item \textbf{Q. 18:} \textbf{[If Q.17 answer is \quotes{yes}]} Approximately, how often do you manually delete these data? \par 
\textbf{Answers:} \begin{inparaitem}[$\circ$] \item \quotes{At least one a day,} \item \quotes{At least once a week,} \item \quotes{At least once a month,} \item \quotes{At least once every 3 months,} \item \quotes{At least once every 6 months,} \item \quotes{At least once a year,} \item \quotes{Less than once a year,} \item \quotes{I never deleted them,} \item \quotes{I do not remember}\end{inparaitem} \vspace{4pt}

\item \textbf{Q. 19:} \textbf{[If Q.13 answer is \quotes{yes}]} Did you know about the Auto-delete feature which allows you to specify how long Google saves the [Web \& App Activity $|$ YouTube History $|$ Location History]\footref{note1} data before they are automatically deleted? (yes/no, elaborate if possible) \par 
\textbf{Answers:} \begin{inparaitem}[$\circ$] \item \quotes{Yes,} \item \quotes{No}\end{inparaitem}
\end{itemize}
\subsection{Auto-delete}
\begin{itemize}
	\item \textbf{Q. 20:} Do you think that the default retention period specified by Google (18 months, i.e. 1.5 years) to save the Web \& App Activity data at Google before auto-deletion is suitable? (yes/no, elaborate if possible) \par 
	\textbf{Answers:} \begin{inparaitem}[$\circ$] \item \quotes{Yes,} \item \quotes{No}\end{inparaitem} 
\vspace{4pt}
	\item\textbf{Q. 21} \textbf{[If Q. 20 answer is \quotes{no}]} What period do you suggest for saving the Web \& App Activity data before Google automatically deletes them? \par 
	\textbf{Answers:} Open 
 \vspace{4pt}
	\item \textbf{Q. 22:} Do you think that the default retention period specified by Google (36 months, i.e. 3 years) to save the YouTube History data is suitable? (yes/no, elaborate if possible) \par 
	\textbf{Answers:} \begin{inparaitem}[$\circ$] \item \quotes{Yes,} \item \quotes{No}\end{inparaitem} 
 \vspace{4pt}
	\item\textbf{Q. 23} \textbf{[If Q. 22 answer is \quotes{no}]} What period do you suggest for saving the YouTube History data at Google before Google automatically deletes them? \par 
	\textbf{Answers:} Open  
 \vspace{4pt}
	\item \textbf{Q. 24:} Do you think that the default retention period specified by Google (18 months, i.e. 1.5 years) to save the Location History data is suitable? (yes/no, elaborate if possible) \par
 	\textbf{Answers:} \begin{inparaitem}[$\circ$] \item \quotes{Yes,} \item \quotes{No}\end{inparaitem} 
  \vspace{4pt}
	\item\textbf{Q. 25} \textbf{[If Q. 24 answer is \quotes{no}]} What period do you suggest for saving the Location History data at Google before Google automatically deletes them? \par 
	\textbf{Answers:} Open 
 \vspace{4pt}
 \end{itemize}
\subsection{Your Data Privacy and Sensitivity}
\begin{itemize}
	\item \textbf{Q. 26:} If Google saves the following data: Web \& App activity, Location History, and YouTube History. Rank them according to the degree of privacy (no. 1 High Privacy, 2 Medium, 3 Low). Note: you can choose more than one item with an equal degree of privacy. \par 	
	\textbf{Answers:} \begin{inparaitem}[$\circ$] \item \quotes{1 (High),} \item \quotes{2 (Medium),} \item \quotes{3 (Low)} \end{inparaitem} 
 \vspace{4pt}
	\item \textbf{Q. 27:} To what extent do you consider your [Web \& App Activity, YouTube History, Location History]\footnote{Asked as 3 separate questions, one question per data type.} data at Google services sensitive? \par 
	\textbf{Answers:} \begin{inparaitem}[$\circ$] \item \quotes{They are all sensitive,} \item \quotes{Some are sensitive,} \item \quotes{All are insensitive,} \item \quotes{Other (*required: please specify)} \end{inparaitem} 
 \vspace{4pt}
 \end{itemize}
\subsection{Acceptability for Saving Your Web \& App Activity}
\begin{itemize}
	\item \textbf{Q. 28:} To what extent would you accept if Google saves, processes, and analyzes your Web \& App activity data, for the purpose of improving the services provided to you, such as speeding up your searches, providing you with better recommendations, and more personalized experiences at Google services? \par 
	\textbf{Answers:} \begin{inparaitem}[$\circ$] \item \quotes{Very acceptable,} \item \quotes{Somewhat acceptable,} \item \quotes{Neutral,} \item \quotes{Somewhat unacceptable,} \item \quotes{Completely unacceptable} \end{inparaitem} 
 \vspace{4pt}
	\item \textbf{Q. 29:} to what extent would you accept if Google saves, processes, and analyzes your Web \& App activity data, for the purpose of displaying ads to you on Google services such as Google search engine or YouTube? \par 
	\textbf{Answers:} \begin{inparaitem}[$\circ$] \item \quotes{Very acceptable,} \item \quotes{Somewhat acceptable,} \item \quotes{Neutral,} \quotes{Somewhat unacceptable,} \item \quotes{Completely unacceptable}  \end{inparaitem} 
 \vspace{4pt}
	\item \textbf{Q. 30:} To what extent would you accept if Google saves, processes, and analyzes your Web \& App activity data, for the purpose of displaying ads to you on non-Google websites and apps that have a partnership with Google (such as newspaper websites and gaming apps)? \par 
	\textbf{Answers:} \begin{inparaitem}[$\circ$] \item \quotes{Very acceptable,} \item \quotes{Somewhat acceptable,} \item \quotes{Neutral,} \item \quotes{Somewhat unacceptable,} \item \quotes{Completely unacceptable} \end{inparaitem} 
\end{itemize}	

\subsection{Privacy Concerns}
\begin{itemize}
	\item \textbf{Q. 31:} Previously, did you have any concerns regarding your privacy while using Google services, such as the Google search engine, Google Maps, or YouTube? \par
	\textbf{Answers:} \begin{inparaitem}[$\circ$] \item \quotes{Yes,} \item \quotes{No,} \item \quotes{Neutral} \end{inparaitem} 
 \vspace{4pt}
	\item \textbf{Q. 32:} \textbf{[If Q. 31 answer is \quotes{yes}]} If your answer is yes, what are these concerns? \par 
	\textbf{Answers:} Open 
 \vspace{4pt}
	\item \textbf{Q. 33:} Have you ever taken any steps to protect your privacy online while using Google services, such as the Google search engine, Google Maps, or YouTube? \par 
	\textbf{Answers:} Open with \begin{inparaitem}[$\circ$] \item \quotes{Yes,} \item \quotes{No}\end{inparaitem}
 \vspace{4pt}
	\item \textbf{Q. 34:} [If Q. 33 answer is \quotes{yes}] If your answer is yes, what are these steps? \par 
	\textbf{Answers:} Open
 \vspace{4pt}
	\item \textbf{Q. 35:} Do you plan to take any additional steps to protect your privacy online while using Google services, such as the Google search engine, Google Maps, or YouTube? \par
	\textbf{Answers:} Open with \begin{inparaitem}[$\circ$] \item \quotes{Yes,} \item \quotes{No}\end{inparaitem}
 \vspace{4pt}
	\item \textbf{Q. 36:} [If Q. 35 answer is \quotes{yes}] If your answer is yes, what are these steps? \par 
	\textbf{Answers:} Open
\end{itemize}

\end{document}